\documentclass[onecolumn,numbers]{els-mrw} 
\usepackage{amsmath,amssymb,amsfonts,amsthm,makeidx,graphicx}
\usepackage{txfonts}
\usepackage{helvet}
\usepackage{graphicx}
\usepackage{hyperref}
\usepackage{color}
\hypersetup{
    pdfnewwindow=true,      
    colorlinks=true,       
    linkcolor=blue,          
    citecolor=blue,        
    filecolor=blue,      
    urlcolor=blue        
}

\begin{document}


\chapter{Neutrino-Nucleus Scattering Cross Sections at Medium Energies}\label{chap9}
\author[1]{Vishvas Pandey}
\address[1]{\orgname{Fermi National Accelerator Laboratory}, \orgaddress{Batavia, Illinois 60510, USA}}
\address[1]{\email{vpandey@fnal.gov}}


\maketitle


\begin{abstract}[Abstract]

The weak interactions of neutrinos with other Standard Model particles are well described within the Standard Model of particle physics. However, modern accelerator-based neutrino experiments employ \emph{nuclei} as targets, where neutrinos interact with bound nucleons, turning a seemingly simple electroweak process into a complex many-body problem in nuclear physics. At the time of writing this \textit{Encyclopedia of Particle Physics} chapter, neutrino--nucleus interactions remain one of the leading sources of systematic uncertainty in accelerator-based neutrino oscillation measurements.

This chapter provides a pedagogical overview of neutrino interactions with nuclei in the \emph{medium-energy} regime, spanning a few hundred~MeV to several~GeV. It introduces the fundamental electroweak formalism, outlines the dominant interaction mechanisms---including quasielastic scattering, resonance production, and deep inelastic scattering---and discusses how nuclear effects such as Fermi motion, nucleon--nucleon correlations, meson--exchange currents, and final-state interactions modify observable cross sections. The chapter also presents a brief survey of the foundational and most widely used theoretical models for neutrino--nucleus cross sections, together with an overview of current and upcoming accelerator-based neutrino oscillation experiments that are shaping the field.

Rather than targeting experts, this chapter serves as a primer for advanced undergraduates, graduate students, and early-career researchers entering the field. It provides a concise foundation for understanding neutrino--nucleus scattering, its relevance to oscillation experiments, and its broader connections to both particle and nuclear physics.

\end{abstract}


\begin{keywords}
    Accelerator Neutrinos\sep Neutrino Interactions\sep Neutrino--Nucleus Scattering\sep Neutrino Cross Sections\sep Nuclear Effects\sep Neutrino Oscillations\sep Electroweak Interactions\sep Medium-Energy Neutrinos
\end{keywords}


\begin{glossary}[Nomenclature]
	\begin{tabular}{@{}lp{34pc}@{}}

    	CC & Charged Current\\
        NC & Neutral Current\\
        QE & Quasielastic Scattering\\
        RES & Resonance Production\\
        SIS & Shallow Inelastic Scattering\\
        DIS & Deep Inelastic Scattering\\
        MEC & Meson--Exchange Currents\\
        SRC & Short--Range Correlations\\
        FSI & Final--State Interactions\\
        LDA & Local Density Approximation\\
        RPA & Random Phase Approximation\\
        IA  & Impulse Approximation\\
        SF  & Spectral Function\\
        $\chi$EFT & Chiral Effective Field Theory\\
        LArTPC & Liquid Argon Time Projection Chamber\\
	\end{tabular}
\end{glossary}


\section*{Objectives}

This chapter in the \textit{Encyclopedia of Particle Physics} provides a pedagogical overview of the 
theoretical and experimental foundations of neutrino--nucleus interactions in the medium-energy regime. 
It is designed as a concise, self-contained reference for readers beginning research or study in this area. 
Specifically, the chapter aims to:

\begin{itemize}
    \item Present the fundamental formalism governing neutrino interactions with nuclear targets, 
    including the electroweak Lagrangian, key kinematic variables, and the definition of differential 
    and total cross sections.
    
    \item Explain the dominant interaction mechanisms across the energy range relevant to 
    accelerator-based neutrino experiments---quasielastic scattering, resonance production, and 
    deep inelastic scattering---and discuss how nuclear effects such as nucleon--nucleon correlations, 
    meson--exchange currents, and final-state interactions modify these processes.
    
     \item Provide an overview of the foundational and most widely used theoretical models, along with current and upcoming accelerator-based neutrino experiments, which together define the contemporary landscape of neutrino--nucleus interaction studies and their impact on precision oscillation physics.
    
    \item Serve as a pedagogical entry point for graduate students, postdoctoral researchers, and 
    physicists from related disciplines seeking a unified understanding of neutrino--nucleus 
    scattering at intermediate energies.
\end{itemize}


\section{Introduction}\label{sec:intro}

Neutrinos are among the most abundant yet least understood particles in the universe. Their extremely weak interactions with matter make them remarkably difficult to study, but also render them powerful messengers of fundamental physics. Understanding neutrinos is essential not only for completing the Standard Model (SM) but also for uncovering phenomena that lie beyond it. The discovery that neutrinos possess mass---a property absent from the SM---already signals new physics, making the detailed exploration of the neutrino sector one of the defining frontiers of modern particle physics~\cite{Huber:2022lpm}.  

This chapter focuses on the \textit{medium-energy regime}---from a few hundred~MeV to several~GeV---which is directly relevant to present and future accelerator-based neutrino oscillation experiments. Accelerator facilities provide a uniquely controlled environment for studying neutrinos: well-characterized beams incident on massive nuclear targets enable high-statistics measurements of cross sections and final-state topologies. Such measurements are indispensable for addressing key open questions: What is the ordering of neutrino masses? Do neutrinos violate CP symmetry? Could additional sterile neutrino states or other exotic interactions exist? The reliability of these answers depends critically on our ability to model and interpret neutrino interactions with the target material in the detector with high precision.  

At first glance, describing neutrino interactions may appear straightforward: one could, in principle, start from the electroweak Lagrangian and compute the relevant amplitudes. In practice, however, the situation is far more intricate. Neutrino scattering from nuclei involves a complex interplay of nuclear structure, many-body correlations, meson-exchange dynamics, and final-state interactions that obscure the underlying weak process. Theoretical frameworks that perform well in one kinematic regime often fail in another, underscoring the challenge of developing a consistent and predictive description across the broad range of energies probed experimentally~\cite{NuSTEC:2017hzk, Balantekin:2022jrq}.  

This chapter is written as a \textit{pedagogical introduction} for graduate students, postdoctoral researchers, and physicists entering the field of neutrino--nucleus interaction physics. It provides a self-contained overview of the theoretical foundations, nuclear modeling methods, and experimental context relevant to neutrino scattering in the medium-energy regime. While familiarity with basic quantum mechanics, statistical physics, and particle physics is assumed, no prior expertise in quantum field theory is required.  

The remainder of this chapter is organized as follows. To set the stage, subsection~\ref{sec:neutrino_sources} introduces accelerator-based neutrino sources, and subsection~\ref{sec:neutrino_oscillation} examines the role of neutrino--nucleus interactions in oscillation physics. Section~\ref{sec:formalism} presents the general formalism for neutrino--nucleus scattering, discussing the dominant interaction channels and the key nuclear effects that influence them. Section~\ref{sec:theory} outlines the principal theoretical frameworks and modeling approaches used to describe these processes. Finally, section~\ref{sec:exp} surveys current and forthcoming experimental programs. Together, these sections provide a cohesive foundation for understanding how neutrinos interact with matter in the medium-energy regime---a cornerstone for achieving precision in future accelerator-based neutrino physics.  

Before diving into the details of neutrino--nucleus scattering physics, let us first review how accelerator-based, medium-energy neutrino beams are produced and characterized, setting the experimental context for the theoretical discussions that follow.

\subsection{Medium-Energy Neutrino Sources}\label{sec:neutrino_sources}

\begin{figure}[t]
    \centering
    \includegraphics[width=1.0\textwidth]{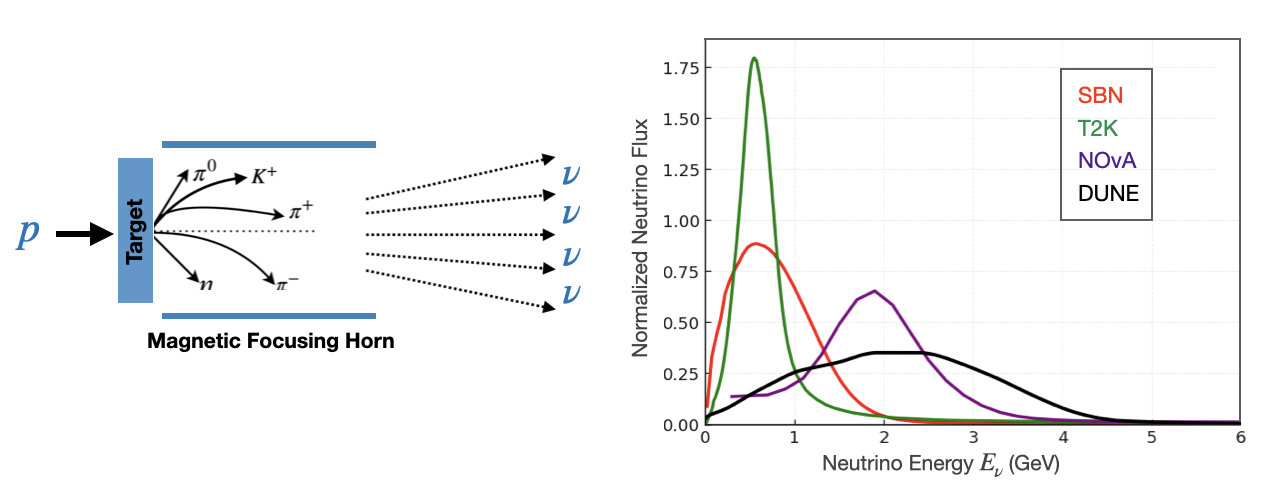}
    \caption{Illustration of neutrino production mechanisms (left) and the resulting medium-energy neutrino fluxes (right) at accelerator-based experiments.}
    \label{fig:neutrino_source}
\end{figure}

Conventional accelerator-based neutrino beams are produced when high-energy protons strike a fixed target, generating cascades of secondary hadrons---primarily charged pions and kaons. These mesons are focused by pulsed magnetic horns and allowed to decay in flight, producing collimated beams of neutrinos. Focusing positively charged hadrons yields a beam dominated by $\nu_\mu$, while reversing the horn polarity selects $\bar{\nu}_\mu$. Despite magnetic focusing, small admixtures of wrong-sign neutrinos and $\nu_e$ components from kaon and muon decays remain~\cite{Kopp:2006ky}.  
Figure~\ref{fig:neutrino_source} illustrates the neutrino production mechanisms (left) and the resulting medium-energy neutrino fluxes (right) at accelerator-based experiments.

Magnetic horns play a central role in controlling both the intensity and charge composition of the beam. By adjusting the horn geometry and current, experiments can tailor the energy spectrum to match the desired oscillation baseline, since the oscillation probability depends on the ratio $L/E_\nu$, as discussed later in Eq.~\ref{eq:osc}. For a pion of energy $E_\pi$ decaying at an angle $\theta$ relative to the beam axis, the neutrino energy is approximately
\begin{equation}
E_\nu = \left[1 - \left(\frac{m_\mu}{m_\pi}\right)^2\right]
\frac{E_\pi}{1 + \gamma^2 \theta^2},
\end{equation}
where $\gamma = E_\pi / m_\pi$ is the Lorentz boost. Forward decays ($\theta \approx 0$) yield the highest-energy neutrinos, whereas larger decay angles produce lower and more sharply peaked spectra.  

The design and optimization of modern accelerator neutrino beams rely critically on precise hadron-production data. Dedicated experiments measure pion and kaon yields from replica targets, providing essential inputs to constrain flux predictions~\cite{NA61SHINE:2025aey}.  
Today’s global accelerator neutrino program is led by Fermilab in the United States and J-PARC in Japan, both of which provide intense, tunable sources of medium-energy neutrinos for oscillation, cross-section, and beyond the Standard Model studies.  

\subsection{Neutrino--Nucleus Interactions and Neutrino Oscillation Physics}\label{sec:neutrino_oscillation}

In accelerator-based neutrino oscillation experiments, neutrino--nucleus interactions constitute one of the dominant sources of systematic uncertainty.  
For a neutrino beam produced in a flavor state $\nu_i$ and detected as $\nu_j$ at the far detector, the observed event rate for a given final-state topology can be expressed schematically as  
\begin{equation}
\mathcal{R}(\nu_i \rightarrow \nu_j)
\propto
\phi_{\nu_i}(E_\nu)
\otimes
\sigma_{\nu_j}(E_\nu)
\otimes
P(\nu_i \rightarrow \nu_j)
\otimes
\epsilon_{\nu_j},
\label{eq:rate}
\end{equation}
where $\phi_{\nu_i}(E_\nu)$ is the neutrino flux, $\sigma_{\nu_j}(E_\nu)$ the interaction cross section, $P(\nu_i \rightarrow \nu_j)$ the oscillation probability, and $\epsilon_{\nu_j}$ the detector efficiency.  

For two-flavor mixing, the oscillation probability is well approximated by  
\begin{equation}
P(\nu_i \rightarrow \nu_j)
\simeq
\sin^2 2\theta\,
\sin^2\!\left(
\frac{\Delta m^2 L}{4E_\nu}
\right),
\label{eq:osc}
\end{equation}
where $\theta$ is the mixing angle, $\Delta m^2$ the mass-squared difference, $E_\nu$ the neutrino energy, and $L$ the baseline.  
In practice, the neutrino energy is reconstructed from the kinematics of visible final-state particles, introducing model-dependent biases due to nuclear effects in neutrino--nucleus scattering.

Near detectors in a typical two-detector (near and far) oscillation setup are designed to constrain these uncertainties.  
However, differences in geometry, acceptance, detector technology, and target composition between the near and far detectors limit the extent to which correlated uncertainties can be canceled.  
Moreover, because the far-detector spectrum is both oscillated and flavor-transformed, the product $\phi(E_\nu)\otimes\sigma(E_\nu)$ cannot be fully factorized from the oscillation probability term.  

Accurate modeling of neutrino--nucleus interactions is therefore essential.  
Processes such as quasielastic scattering, multinucleon excitations, resonance production, and deep inelastic scattering contribute differently to reconstructed energy distributions and are governed by distinct nuclear dynamics, as discussed in Sec.~\ref{sec:formalism}.  
If these effects are not properly modeled, they can introduce biases in the extraction of oscillation parameters such as $\delta_{CP}$ or the neutrino mass ordering.  
Achieving the percent-level systematic precision required for next-generation long-baseline experiments demands improved theoretical modeling, comprehensive cross-section measurements, and consistent treatment of nuclear effects across all interaction channels~\cite{NuSTEC:2017hzk, Balantekin:2022jrq}. To understand how these nuclear effects enter neutrino oscillation analyses, we now turn to the theoretical foundations of neutrino--nucleus scattering, outlining the basic formalism, kinematics, and cross-section structure that form the basis of modern neutrino-nucleus scattering cross section physics.


\section{General Neutrino--Nucleus Scattering Cross Section Formalism}\label{sec:formalism}

At medium energies, where the resolution is coarser, neutrino interactions reveal collective nuclear properties---the overall size, density, and the spatial distributions of protons and neutrons, as well as the response of the nucleus to weak currents and excitations.  
At higher energies, neutrinos probe the substructure of the nucleon, accessing the parton distributions of quarks and antiquarks as a function of their momentum fraction.  
Unlike electromagnetic scattering, weak interactions are sensitive to both vector and axial currents, providing unique access to spin, isospin, and flavor structure. However, the extremely weak coupling and broad energy spectra of neutrinos also require large, high-precision detectors and intense accelerator beams.  

Within the Standard Model (SM), neutrinos are spin-$\tfrac{1}{2}$ leptons that interact exclusively via the weak force.  
They carry neither electric charge nor color and are therefore singlets under the $SU(3)_C \times U(1)_{\text{EM}}$ subgroup of the electroweak gauge group.  
Each neutrino flavor appears as the upper component of a left-handed lepton doublet,
\begin{equation}
L_\ell =
\begin{pmatrix}
\nu_\ell \\
\ell
\end{pmatrix}_L ,
\label{eq:lepton_doublet}
\end{equation}
where $\ell = e,\,\mu,\,\tau$ denotes the corresponding charged lepton. Gauge invariance under $SU(2)_L$ fixes the structure of the weak-interaction Lagrangians.  
Charged-current (CC) interactions couple neutrinos to their charged partners through exchange of $W^\pm$ bosons, while neutral-current (NC) interactions, mediated by the $Z^0$, act among neutrinos and other fermions without changing electric charge:
\begin{align}
\mathcal{L}_{\text{CC}} &=
- \frac{g}{\sqrt{2}}
\sum_{\ell}
\bar{\nu}_{\ell L}\gamma^{\mu}\ell_L^-\,W_{\mu}^{+} + \text{h.c.}, \\
\mathcal{L}_{\text{NC}} &=
- \frac{g}{2\cos\theta_W}
\sum_{\ell}
\bar{\nu}_{\ell L}\gamma^{\mu}\nu_{\ell L}\,Z_{\mu}^{0},
\label{eq:weak_lagrangians}
\end{align}
where $g$ is the $SU(2)_L$ coupling constant and $\theta_W$ is the weak mixing (Weinberg) angle.  
After spontaneous symmetry breaking, these interactions govern all weak processes involving neutrinos, from $\beta$ decay and coherent elastic scattering to high-energy neutrino reactions.

For neutrino--nucleus scattering, the relevant interactions arise from the low-energy limit of these Lagrangians, where the momentum transfer is much smaller than the $W$ or $Z$ boson masses.  
In this regime, the heavy boson propagators can be integrated out, leading to an effective four-fermion Fermi interaction characterized by the constant $G_F$.  
The remainder of this section develops the formal connection between this effective weak interaction and the experimentally measurable neutrino--nucleus cross sections, providing a pedagogical bridge between the electroweak theory and nuclear-scale observables.

The total cross section $\sigma$ quantifies the probability that an incoming neutrino interacts with a target, depending on both the intrinsic interaction dynamics and the external kinematics such as the incident energy.  
Differential cross sections, for example $d\sigma/d\Omega$, specify the probability that the scattered lepton emerges within a given solid angle $\Omega$.   In what follows, we present the general formalism originally developed for polarized electron scattering and later extended to weak interactions, leading to a unified description of electromagnetic and weak lepton--nucleus processes.  
The discussion below focuses primarily on charged current quasielastic scattering, shown schematically in Fig.~\ref{fig:neutrino_interaction_feynmann}(a), but the same structure applies to other channels discussed in Sec.~\ref{sec:interaction_channels}.

Consider a charged–current process in which an incoming neutrino scatters off a target nucleus $A$, producing a charged lepton $\ell$ in the final state.  
The outgoing lepton is detected, while the residual nuclear system is left unobserved.  
The incoming and outgoing lepton four–momenta are denoted
\[
k_i = (\epsilon_i, \vec{k}_i), \qquad
k_f = (\epsilon_f, \vec{k}_f),
\]
and the four–momentum transferred from the neutrino to the nuclear system is
\begin{equation}
q^\mu = (\omega, \vec{q}), \qquad
\omega = \epsilon_i - \epsilon_f, \qquad
\vec{q} = \vec{k}_i - \vec{k}_f.
\end{equation}
The invariant magnitude of the momentum transfer is
\[
Q^2 = \vec{q}^{\,2} - \omega^2,
\]
and another commonly used kinematic variable is the Bjorken scaling variable,
\begin{equation}
x = \frac{Q^2}{2 M_N \omega},
\label{eq:bjorken-x}
\end{equation}
where $M_N$ is the mass of the initial nucleon.  
The invariant mass of the final state hadronic system, $W$, is defined as
\begin{equation}
W = \sqrt{(M_N + \omega)^2 - \vec{q}^{\,2}}
   = \sqrt{M_N^2 + 2 M_N \omega - Q^2}.
\label{eq:invariant-mass}
\end{equation}

In the laboratory frame, the $z$–axis is conventionally chosen along $\vec{q}$, defining the scattering plane.  
For charged–current interactions mediated by $W^\pm$ bosons, the propagator can be approximated by
\begin{equation}
P^{W}_{\mu\nu} \simeq \frac{i\, g_{\mu\nu}}{M_W^2},
\label{eq:W-prop}
\end{equation}
which is valid when the energy transfer $\omega$ is far below the $W$ boson mass ($M_W \simeq 80~\text{GeV}$).  

The differential cross section follows from the invariant transition amplitude, which in the Bjorken–Drell convention takes the general form
\[
\mathcal{M}_{fi}^{W} \propto \langle f | \hat{J}_{\text{lep}}^\mu P^W_{\mu\nu} \hat{J}_{\text{nuc}}^\nu | i \rangle ,
\]
where a delta function in the full expression enforces four–momentum conservation.  
Applying the low–energy limit of the weak interaction, the invariant matrix element for CC scattering becomes
\begin{equation}
\mathcal{M}_{fi}^{W}
= -\,i\,\frac{G_F}{\sqrt{2}}\,\cos\theta_c\,
\mathcal{J}^{\text{lep}}_{\nu}(q)\,
\mathcal{J}^{\nu}_{\text{nuc}}(q),
\label{eq:Mfi-W}
\end{equation}
where $\theta_c$ is the Cabibbo angle, $G_F$ is the Fermi coupling constant, and $\mathcal{J}^{\text{lep}}_\nu$ and $\mathcal{J}^{\nu}_{\text{nuc}}$ denote the leptonic and nuclear current matrix elements, respectively.  
The weak couplings are related by
\begin{equation}
\frac{G_F}{\sqrt{2}} = \frac{g^2}{8 M_W^2},
\label{eq:weak-relations}
\end{equation}
linking the effective four–fermion description to the underlying electroweak theory.

The leptonic and nuclear currents are defined as
\begin{align}
\mathcal{J}^{\text{lep}}_{\mu}(q)
&\equiv \bar{u}(k_f, s_f)\, \hat{J}^{\text{lep}}_\mu\, u(k_i, s_i)
= \bar{u}(k_f, s_f)\, \gamma_\mu (1 + h \gamma^5)\, u(k_i, s_i), 
\label{eq:lep-current} \\
\mathcal{J}^{\text{nuc}}_{\mu}(q)
&\equiv \langle \Psi_f |\, \hat{J}^{\text{nuc}}_\mu(q)\, | \Psi_i \rangle ,
\label{eq:nuc-current}
\end{align}
where $u(k_i, s_i)$ and $\bar{u}(k_f, s_f)$ are the Dirac spinors of the incoming and outgoing leptons, and $|\Psi_i\rangle$, $|\Psi_f\rangle$ represent the initial and final nuclear many–body states.  
The operators $\hat{J}^{\text{lep}}_\mu$ and $\hat{J}^{\text{nuc}}_\mu(q)$ are the leptonic and nuclear current operators in momentum space.  

The leptonic current is fully determined by field theory, with the helicity parameter $h$ taking values $h=+1$ for neutrinos and $h=-1$ for antineutrinos, reflecting the $V\!-\!A$ structure of the weak interaction.  
By convention, the factor $\tfrac{1}{2}$ from the projection operator $(1 + h\gamma^5)/2$ is absorbed into the weak coupling constant.  
The nuclear current, in contrast, is far more complex: it requires evaluating many–body matrix elements between correlated nuclear states and therefore encapsulates the full dynamical content of the nuclear response.

Squaring the invariant matrix element in Eq.~\eqref{eq:Mfi-W} and summing (averaging) over the initial and final spin states yields
\begin{equation}
\sum_{if}\big|\mathcal{M}_{fi}^{W}\big|^2
= \frac{G_F^2}{2}\,\cos^2\theta_c\, L_{\mu\nu} W^{\mu\nu},
\label{eq:Mfi-W-sq}
\end{equation}
where $L_{\mu\nu}$ and $W^{\mu\nu}$ denote the leptonic and hadronic tensors, respectively.  
Each tensor is a bilinear combination of its corresponding current and encapsulates the leptonic and nuclear dynamics of the process.

The \textit{leptonic tensor} is defined as
\begin{equation}
L_{\mu\nu} = 
\sum_{if} 
\big[ \mathcal{J}^{\text{lep}}_{\mu}(q) \big]^{\dagger} 
\mathcal{J}^{\text{lep}}_{\nu}(q),
\label{eq:Lmunu}
\end{equation}
which evaluates to
\begin{equation}
L_{\mu\nu} =
\frac{2}{m_i m_f}
\left(
k_{i,\mu} k_{f,\nu} + k_{f,\mu} k_{i,\nu}
- g_{\mu\nu}\, k_i \!\cdot\! k_f
+ g_{\mu\nu}\, m_i m_f
- i\,h\,\epsilon_{\mu\nu\alpha\beta}\, k_i^{\alpha} k_f^{\beta}
\right),
\label{eq:Lmunu-expanded}
\end{equation}
where $h$ denotes the lepton helicity.  
For neutrino and antineutrino scattering, the initial lepton mass is negligible ($m_i \approx 0$), so the fourth term can be safely omitted.  
The antisymmetric part of $L_{\mu\nu}$, proportional to $\epsilon_{\mu\nu\alpha\beta}$, distinguishes between $\nu$ and $\bar{\nu}$ interactions and gives rise to parity-violating interference terms.

The \textit{hadronic tensor} is defined analogously by summing and averaging over the nuclear spin projections:
\begin{equation}
W^{\mu\nu} =
\frac{1}{2J_i + 1}
\sum_{M_i}
\sum_{J_R, M_R}
\sum_{m_{s_N}}
\big[ J^{\mu}_{\text{nuc}}(q) \big]^{\dagger}
J^{\nu}_{\text{nuc}}(q),
\label{eq:Wmunu}
\end{equation}
where $(J_i, M_i)$ are the spin quantum numbers of the initial nucleus, $(J_R, M_R)$ those of the residual nucleus, and $m_{s_N}$ denotes the spin projection of the ejected nucleon.  
For spin–zero nuclei, the factor $(2J_i + 1)^{-1}$ equals unity and the sum over $M_i$ can be omitted.  
Unlike the purely kinematic leptonic tensor, the hadronic tensor encodes the full complexity of the nuclear response, involving many-body operators evaluated between correlated nuclear states.

With the momentum transfer $\vec{q}$ defining the $\hat{z}$–axis, the contraction of the leptonic and hadronic tensors yields
\begin{equation}
L_{\mu\nu} W^{\mu\nu}
= \frac{2\,\epsilon_i \epsilon_f}{m_i m_f}
\Big(
v_{CC} W_{CC}
+ v_{CL} W_{CL}
+ v_{LL} W_{LL}
+ v_{T} W_{T}
\pm v_{T'} W_{T'}
\Big),
\label{eq:LmunuWmunu}
\end{equation}
where the coefficients $v_i$ are purely kinematic factors depending only on lepton variables, and the $W_i$ represent nuclear response functions that encapsulate target dynamics.  
The $\pm$ sign corresponds to neutrino ($+$) and antineutrino ($-$) scattering.  

This decomposition clearly separates lepton kinematics from nuclear structure, forming the basis for calculating both exclusive and inclusive neutrino--nucleus cross sections. The response functions $W_i$ are directly related to specific components of the nuclear current.  
Their evaluation depends on the underlying nuclear model, and the principal theoretical approaches used to compute them will be discussed later in Sec.~\ref{sec:theory}.

The double–differential cross section can now be written as
\begin{equation}
\frac{d\sigma}{d\omega\, d\Omega}
=\left(\frac{G_F \cos\theta_c}{2\pi}\right)^2
\epsilon_f\, k_f\,
\Big(
v_{CC} W_{CC}
+ v_{CL} W_{CL}
+ v_{LL} W_{LL}
+ v_{T} W_{T}
\pm v_{T'} W_{T'}
\Big),
\label{eq:diff_xs}
\end{equation}
where $\theta_c$ is the Cabibbo angle, $\Omega$ represents the solid angle of the outgoing lepton.  
The nuclear responses $W_i$ depend only on the energy and momentum transfer $(\omega, q)$, while the dependence on the lepton kinematics enters entirely through the kinematic coefficients $v_i$.  
For plane–wave leptons, these factors take the explicit forms~\cite{Walecka2004, Donnelly:1985ry}:
\begin{align}
v_{CC} &= 1 + \frac{k_f}{\epsilon_f}\cos\theta_f, \nonumber\\
v_{CL} &= -\left(\frac{\omega}{q}\,v_{CC} + \frac{m_f^2}{\epsilon_f q}\right), \nonumber\\
v_{LL} &= v_{CC} - \frac{2\,\epsilon_i \epsilon_f}{q^2}\left(\frac{k_f}{\epsilon_f}\right)^2\sin^2\theta_f, \label{eq:vi_factors}\\
v_{T}  &= 2 - v_{CC} + \frac{\epsilon_i \epsilon_f}{q^2}\left(\frac{k_f}{\epsilon_f}\right)^2\sin^2\theta_f, \nonumber\\
v_{T'} &= \frac{\epsilon_i + \epsilon_f}{q}\,(2 - v_{CC}) - \frac{m_f^2}{\epsilon_f q}. \nonumber
\end{align}

To relate these response functions to the underlying nuclear structure, it is convenient to decompose the nuclear current into its spherical components in the reference frame where $\vec{q}$ defines the $z$–axis.  
The vector current is expanded in a basis of irreducible tensor operators, constructed using vector spherical harmonics $\vec{\mathcal{Y}}^{M}_{J(L,1)}$.  
Following Refs.~\cite{Walecka2004, OConnell:1972edu}, the relevant operators are
\begin{align}
\mathcal{M}_{J} &= \int d\vec{r}\, [j_J(qr) Y_{JM}(\Omega_r)]\, \mathcal{J}^0(\vec{r}), \label{eq:MJ}\\
\mathcal{L}_{J} &= \frac{i}{q} \int d\vec{r}\, \big[\vec{\nabla} (j_J(qr) Y_{JM}(\Omega_r))\big] \cdot \vec{\mathcal{J}}(\vec{r}), \label{eq:LJ}\\
\mathcal{T}^{el}_{J} &= \frac{1}{q} \int d\vec{r}\, \big[\vec{\nabla} \times j_J(qr)\, \vec{\mathcal{Y}}^{M}_{J(J,1)}(\Omega_r)\big] \cdot \vec{\mathcal{J}}(\vec{r}), \label{eq:Tel}\\
\mathcal{T}^{mag}_{J} &= \int d\vec{r}\, [j_J(qr)\, \vec{\mathcal{Y}}^{M}_{J(J,1)}(\Omega_r)] \cdot \vec{\mathcal{J}}(\vec{r}). \label{eq:Tmag}
\end{align}

These correspond to the Coulomb ($\mathcal{M}_J$), longitudinal ($\mathcal{L}_J$), transverse electric ($\mathcal{T}^{el}_J$), and transverse magnetic ($\mathcal{T}^{mag}_J$) operators.  
As the matrix element is now written as a function of irreducible tensor operators evaluated between good angular momentum eigenstates, the Wigner–Eckart theorem can be used to write the expectation value of these operators as a product of a $3j$-symbol and a reduced matrix element 
involving only the radial wavefunctions. 

Squaring the transition amplitude and summing over unobserved initial and final state quantum numbers then leads to the following expressions for the nuclear response functions in terms of these reduced matrix elements:
\begin{align}
W_{CC} &= \sum_{J \ge 0} \sum_{J_f, J_i} 
\big|\langle J_f \| \mathcal{M}_J \| J_i \rangle\big|^2, \label{eq:RCC}\\
W_{LL} &= \sum_{J \ge 0} \sum_{J_f, J_i} 
\big|\langle J_f \| \mathcal{L}_J \| J_i \rangle\big|^2, \label{eq:RLL}\\
W_{CL} &= \sum_{J \ge 0} \sum_{J_f, J_i} 
2\, \mathrm{Re}\!\left[\langle J_f \| \mathcal{M}_J \| J_i \rangle 
\langle J_f \| \mathcal{L}_J \| J_i \rangle^* \right], \label{eq:RCL}\\
W_{T} &= \sum_{J \ge 1} \sum_{J_f, J_i} 
\Big(\big|\langle J_f \| \mathcal{T}^{el}_J \| J_i \rangle\big|^2 
+ \big|\langle J_f \| \mathcal{T}^{mag}_J \| J_i \rangle\big|^2 \Big), \label{eq:RT}\\
W_{T'} &= \sum_{J \ge 1} \sum_{J_f, J_i} 
2\, \mathrm{Re}\!\left[\langle J_f \| \mathcal{T}^{el}_J \| J_i \rangle 
\langle J_f \| \mathcal{T}^{mag}_J \| J_i \rangle^* \right], \label{eq:RTprime}
\end{align}
where the sums over $J_i$ and $J_f$ are constrained by angular–momentum coupling.  
This formulation makes explicit which components of the nuclear current contribute to each response, thereby linking measurable observables to the microscopic nuclear dynamics probed by the neutrino.

\begin{figure}[t]
    \centering
    \includegraphics[width=1.0\textwidth]{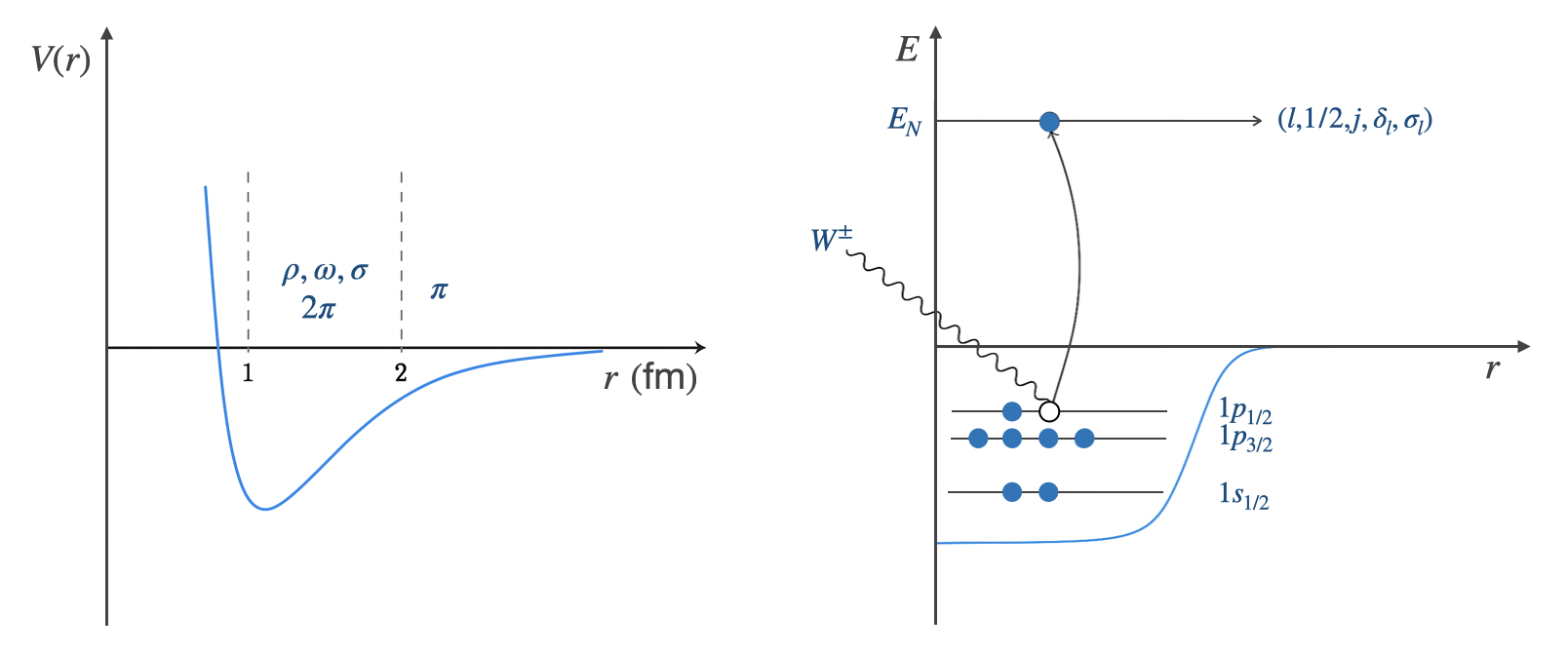}
    \caption{(Left) Radial dependence of the nucleon--nucleon potential, illustrating its long-range attraction and short-range repulsion. (Right) Schematic representation of a one-nucleon knockout in a shell-model picture, where the residual nucleus is left in a one-hole state.}
    \label{fig:potential_shells}
\end{figure}

Microscopic nuclear many-body approaches aim to describe the structure and dynamics of atomic nuclei in terms of the underlying interactions among protons and neutrons, which are treated as the fundamental degrees of freedom.  
In this framework, the nucleus is modeled as a collection of nonrelativistic point-like nucleons whose dynamics are governed by the Hamiltonian
\begin{equation}
H = \sum_i K_i + \sum_{i<j} v_{ij} + \sum_{i<j<k} V_{ijk},
\end{equation}
where $K_i$ denotes the single-nucleon kinetic energy, $v_{ij}$ the two-nucleon (NN) potential, and $V_{ijk}$ the three-nucleon (3N) interaction term.  
Higher-order many-body forces beyond 3N are typically neglected.  

A complete determination of the NN interaction directly from Quantum Chromodynamics (QCD) remains one of the major challenges in modern nuclear theory.  
While lattice QCD has made significant progress toward this goal, present calculations remain approximate and rely on simplifying assumptions.  
In practice, phenomenological potentials—constructed to reproduce NN scattering data and deuteron properties—continue to provide the practical foundation for nuclear structure and reaction studies.

The NN potential has the general form shown in Fig.~\ref{fig:potential_shells}~(left), and can be broadly characterized by three spatial regimes:
\begin{itemize}
    \item \textbf{Long-range region} ($r \gtrsim 2~\text{fm}$): dominated by one-pion exchange, providing the primary attractive force that binds nucleons together.  
    \item \textbf{Intermediate-range region} ($1~\text{fm} \lesssim r \lesssim 2~\text{fm}$): governed by the exchange of multiple mesons, particularly $\sigma$, $\rho$, and $\omega$, introducing both attraction and spin--isospin dependence.  
    \item \textbf{Short-range region} ($r \lesssim 1~\text{fm}$): characterized by a strong repulsive core, essential for explaining the stability and saturation of nuclear matter.  
\end{itemize}
This short-range repulsion prevents nucleons from collapsing into each other and plays a critical role in determining the equation of state of dense nuclear matter, with implications for neutron stars and core-collapse supernovae.  
Traditionally, phenomenological NN interactions were built by combining long-range one-pion exchange with intermediate- and short-range components tuned to NN scattering data~\cite{Wiringa:1994wb, Machleidt:2000ge}.  
More recently, chiral effective field theory ($\chi$EFT) has provided a systematically improvable framework linking these interactions to the symmetries of QCD, yielding a consistent description of both nuclear forces and their electroweak couplings~\cite{Weinberg:1990rz, Ordonez:1995rz, Epelbaum:2008ga, Machleidt:2011zz}.

The formalism developed in this section provides the theoretical backbone for describing neutrino--nucleus interactions.  
By factorizing the cross section into leptonic kinematic coefficients $v_i$ and nuclear response functions $W_i$, it explicitly shows how nuclear dynamics enter experimentally measurable observables.  
In practice, the predictive power of this framework depends critically on how accurately the nuclear response functions are computed—whether through microscopic many-body methods, effective models, or phenomenological fits.  
Section~\ref{sec:theory} will examine how different theoretical approaches implement these calculations, connecting the formal structure developed here to practical modeling frameworks used in contemporary neutrino physics. With the general scattering formalism in place, we will now examine the dominant reaction mechanisms that govern neutrino interactions in the medium-energy domain.  


\subsection{Dominant Interaction Channels}\label{sec:interaction_channels}  

\begin{figure}[t]
    \centering
    \includegraphics[width=1.0\textwidth]{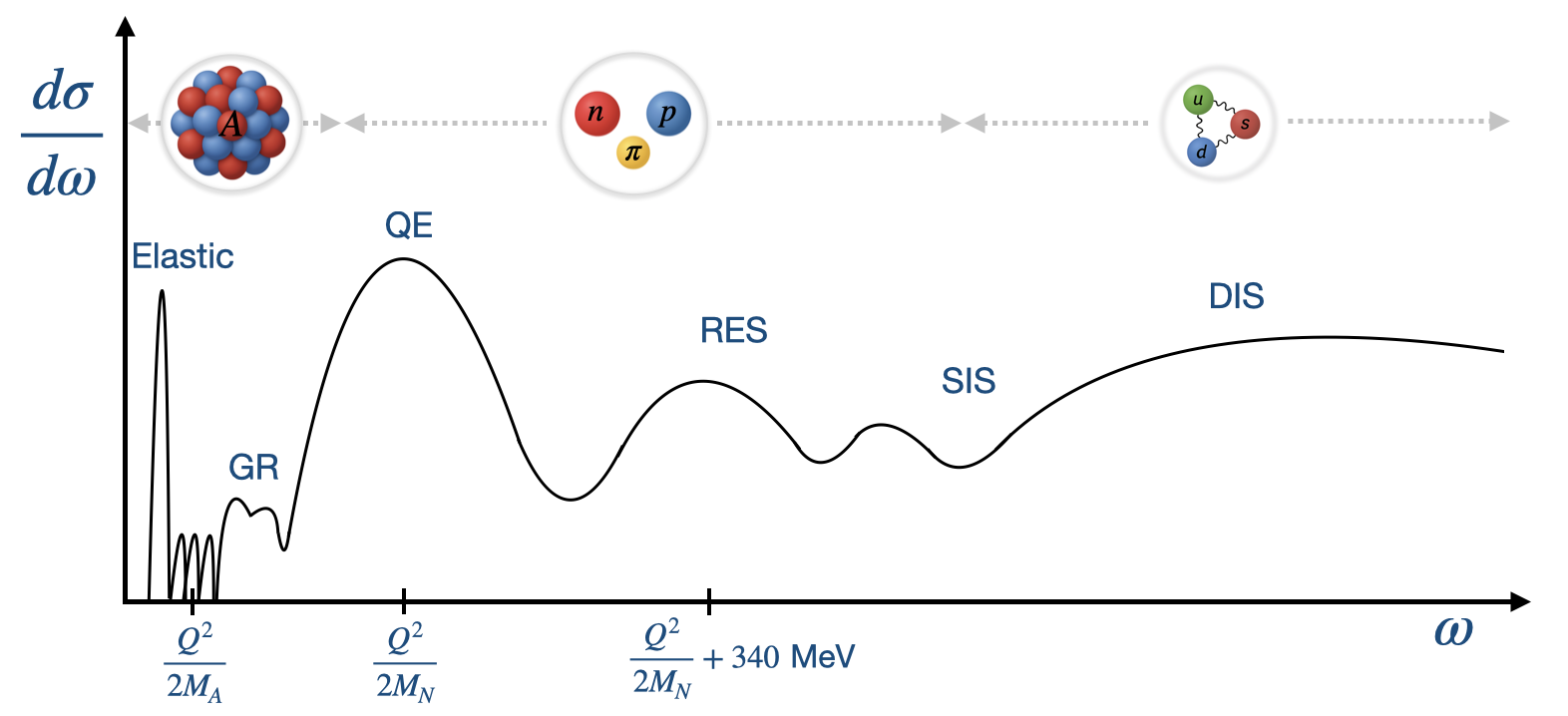}
    \caption{Schematic representation of neutrino–nucleus scattering as a function of energy transfer to the nucleus, highlighting the dominant interaction processes. The top of the schematic illustrates the corresponding evolution of effective degrees of freedom.}
    \label{fig:neutrino_interaction}
\end{figure}

Depending on the incident neutrino energy and the magnitude of the energy and momentum transfer $(\omega,\,q)$, different physical mechanisms dominate. Figure~\ref{fig:neutrino_interaction} illustrates this evolution schematically, showing how the nuclear response varies with energy transfer to the nucleus. At the lowest energies, neutrinos probe elastic scattering and low-lying nuclear excitations, including collective modes such as giant resonances, where the relevant degrees of freedom are entire nuclei and underlying nuclear structure.  
With increasing energy transfer, quasielastic scattering becomes dominant corresponding to the knockout of individual nucleons, followed by resonance excitations and meson production -- where nucleons and pions become the active degrees of freedom.  
Finally, at sufficiently high energy transfer, deep inelastic scattering occurs, and the neutrino interaction probes quarks inside the nucleon. These regimes are not sharply separated but instead form a smooth transition from collective nuclear behavior to hadronic and eventually partonic degrees of freedom. The following subsections examine the dominant interaction channels in the medium-energy domain, outlining the underlying reaction mechanisms, and characteristic kinematics.

\subsubsection{Quasielastic Scattering}

At very low momentum transfers, neutrinos may scatter coherently from the entire nucleus (known as the CE$\nu$NS process), while slightly higher energies excite collective modes such as \textit{giant resonances}~\cite{Pandey:2023arh}.  
At low to intermediate energy transfers, the neutrino begins to resolve individual nucleons within the nucleus, giving rise to \textit{quasielastic} (QE) scattering.  
The QE region thus marks the transition from collective nuclear excitations to single-nucleon dynamics.  
In this regime, the reaction proceeds predominantly through the knockout of a single nucleon via the weak charged current,
\[
\nu_\mu + A \rightarrow \mu^- + p + (A-1),
\]
producing a charged lepton and an outgoing nucleon, as illustrated in Fig.~\ref{fig:neutrino_interaction_feynmann} (a).  
In Fig.~\ref{fig:potential_shells} (right), a one-nucleon knockout process is shown in a shell-model picture: the exchanged boson ejects a proton from the $1p_{1/2}$ shell, leaving a one-hole state in the residual nucleus.  
This process dominates the total cross section below about 1~GeV and forms the cornerstone of energy reconstruction in accelerator-based oscillation experiments.  

Beyond pure single-nucleon knockout, multinucleon mechanisms such as two-particle–two-hole (2p–2h) excitations arising from meson-exchange currents (MEC) or short-range correlations (SRC) contribute significantly.  
In these processes, the neutrino interacts with a correlated nucleon pair, leading to the emission of two nucleons.  
Modern analyses adopt the \textit{CC0$\pi$} category—charged-current events featuring a visible lepton and no pions—to encompass both 1p–1h and 2p–2h topologies.  

Quasielastic scattering provides a sensitive probe of nuclear structure and dynamics.  
Its cross section depends not only on the kinematic variables—energy and momentum transfer—but also on nuclear properties such as mean-field effects, correlations, and final-state interactions.  
Accurate modeling of these ingredients is essential, as they alter the QE peak’s shape, shift reconstructed neutrino energies, and affect the multiplicities and angular spectra of emitted nucleons.  
Recent measurements across multiple experiments have provided high-statistics QE data on various nuclear targets, using both lepton and hadron kinematics to constrain nuclear models.  
Observables such as proton multiplicities, transverse kinematic imbalances, and nucleon angular distributions have demonstrated that even this “simplest” interaction channel is profoundly shaped by complex many-body nuclear dynamics.

\subsubsection{Resonance Production}

\begin{figure}[t]
    \centering
    \includegraphics[width=1.0\textwidth]{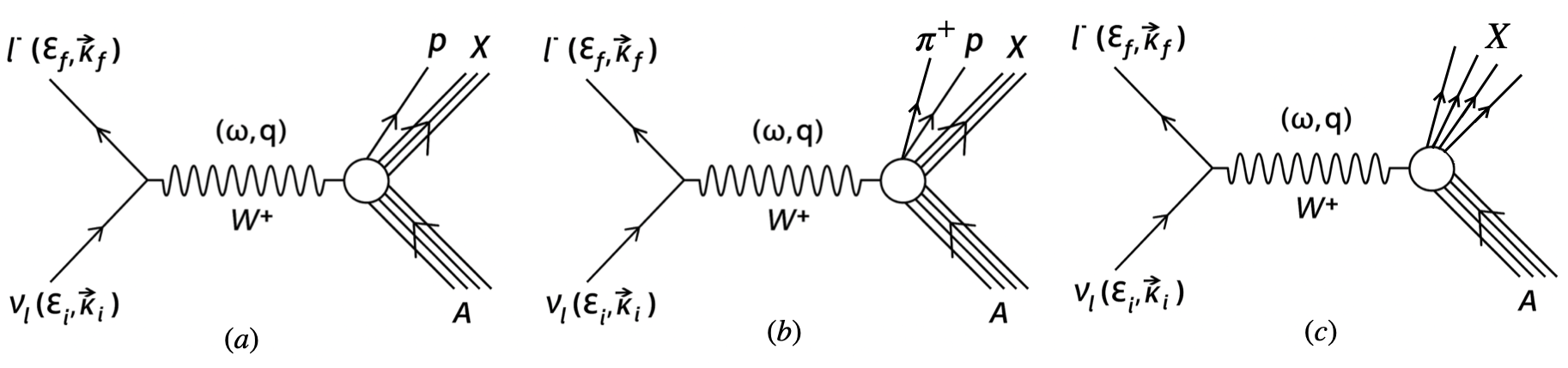}
    \caption{Diagrammatic representation of dominant neutrino–nucleus scattering channels: (a) quasielastic scattering, (b) resonance production, and (c) deep-inelastic scattering.}
    \label{fig:neutrino_interaction_feynmann}
\end{figure}

As the energy transfer increases beyond the quasielastic regime, neutrinos can excite bound nucleons to baryon resonances, leading to meson production in the final state, an example is shown in Fig.~\ref{fig:neutrino_interaction_feynmann}~(b).  
The dominant reaction channels include $\pi N$, $\pi\pi N$, $\eta N$, and $K N$, as well as associated strangeness production ($K Y$ with $Y=\Lambda,\Sigma,\ldots$).  
In the few-GeV range relevant for accelerator neutrinos, single-pion production dominates.  
Despite significant progress in the last decade, resonance production remains one of the least constrained components of neutrino–nucleus scattering at medium energies.

At low hadronic invariant masses $W$, the process is governed by the excitation of the $\Delta(1232)$ resonance, which subsequently decays to a nucleon and a pion.  
As the invariant mass $W$ increases, higher resonances such as the $P_{11}(1440)$, $D_{13}(1520)$, and $S_{11}(1535)$ contribute, opening multi-pion and heavier-meson production channels.  
Roughly twenty baryon resonances populate the region $W < 2~\text{GeV}$, each characterized by distinct spin, parity, and isospin quantum numbers.  

In the nuclear medium, resonance properties—such as mass, width, and decay branching ratios—are modified by interactions with surrounding nucleons.  
Produced pions can undergo strong final-state interactions (FSI): elastic scattering, charge exchange (e.g., $\pi^- p \rightarrow \pi^0 n$), or absorption ($\pi NN \rightarrow NN$).  
Pion absorption, in particular, effectively removes the pion from the final state, making resonance-induced events experimentally indistinguishable from QE-like topologies.  
For example, in the process
\[
\nu_\mu + A \rightarrow \mu^- + \Delta^+ + \ldots,
\]
\[
\Delta^+ \rightarrow p + \pi^0 \ \ \text{or}\ \ n + \pi^+,
\]
the emitted pion may be reabsorbed ($\pi NN \!\rightarrow\! NN$) or scatter within the nucleus before escaping, making them experimentally indistinguishable from QE-like interactions.  
In-medium modifications of resonance masses and widths, especially for the $\Delta(1232)$, are also closely connected to meson-exchange currents and short-range correlations, underscoring the need for consistent nuclear modeling to avoid double counting.

Theoretical descriptions of resonance production must include both resonant and nonresonant amplitudes.  
While resonant processes proceed through intermediate baryon excitations, nonresonant contributions—arising from background Born terms and meson-exchange mechanisms—interfere with them and must be treated consistently.  
Vector transition form factors are constrained by electron- and pion-scattering data, while axial ones rely mainly on legacy bubble-chamber measurements on hydrogen and deuterium.  
Near threshold, chiral perturbation theory provides a systematic expansion, but at higher energies phenomenological models are required.  
The Partial Conservation of the Axial Current (PCAC) hypothesis and the Goldberger–Treiman relation link axial currents at $Q^2 = 0$ to $\pi N$ scattering, though the finite-$Q^2$ behavior remains uncertain.  

Together, these effects make resonance production a particularly rich yet challenging regime in neutrino–nucleus scattering—one that bridges the transition from single-nucleon dynamics to the onset of partonic degrees of freedom, and remains an active frontier for both theoretical and experimental study.

\subsubsection{Shallow and Deep Inelastic Scattering}

At neutrino energies of a few~GeV and above, the interaction dynamics gradually transition from hadronic to partonic degrees of freedom.  
At lower invariant masses $W$, the process can be described in terms of nucleons and resonances, while at higher momentum transfers, the neutrino resolves the quark and gluon constituents of the nucleon.  
The intermediate domain bridging these descriptions is known as the \emph{shallow inelastic scattering} (SIS) region.  
Its boundaries are phenomenological and overlap with both the resonance tail and the onset of deep inelastic scattering (DIS).  
In this transitional region, nonresonant meson production, higher-twist effects, and multi-quark correlations coexist, making a consistent theoretical treatment particularly challenging.

At higher momentum transfers and invariant masses, neutrinos interact directly with quarks inside nucleons through $W^\pm$ or $Z^0$ exchange.  
This process defines the deep inelastic scattering (DIS) regime, where the inclusive charged-current (CC) reaction is
\begin{equation}
\nu_\mu + A \;\rightarrow\; \mu^- \ + X.
\label{eq:DIS-reaction}
\end{equation}
Here, $X$ denotes a hadronic jet formed as the struck quark hadronized, as shown in Fig.~\ref{fig:neutrino_interaction_feynmann}~(c).
In the DIS limit, the cross section factorizes into leptonic and hadronic tensors, with the latter expressed in terms of structure functions $F_i(x,Q^2)$ that depend on the Bjorken scaling variable $x = Q^2/(2 M_N \omega)$ and encode the parton distribution functions (PDFs). In nuclei, these structure functions are modified by several well-known nuclear effects:  
\textit{shadowing} at low $x$ (from coherent multi-nucleon scattering),  
\textit{antishadowing} at intermediate $x$,  
and the \textit{EMC effect} at moderate $x$ (reflecting modifications of bound-nucleon structure).  

The shallow inelastic scattering region, situated between the resonance and DIS regimes, remains a focal point of modern modeling.  
Here, resonant, nonresonant, and partonic mechanisms overlap and interfere.  
Target-mass corrections, higher-twist contributions, and hadronization effects all influence the cross section, while the scarcity of dedicated experimental data continues to limit direct validation of theoretical descriptions.  
Accurate modeling of this transition is essential for oscillation experiments operating in the few-GeV range, such as DUNE, where these processes constitute a significant fraction of observed events.

A unifying concept across this transition is the \textit{quark–hadron duality}~\cite{Bloom:1970xb,Bloom:1971ye}.  
It encapsulates the observation that resonance structure functions at low $Q^2$, when averaged over suitable kinematic intervals, reproduce the scaling curves observed in DIS at high $Q^2$.  
In other words, the quark-level and hadronic descriptions of lepton–nucleon scattering are closely connected when viewed inclusively.  
While duality has been confirmed with high precision in charged-lepton scattering, its validity in neutrino interactions is less certain, primarily because axial and vector currents contribute differently to proton and neutron responses.  
Consequently, duality is expected to hold only approximately for isoscalar targets, and its quantitative realization in neutrino event generators remains an active area of study.

At sufficiently high neutrino energies ($E_\nu \!\gtrsim\! 10$~GeV), the DIS formalism fully applies: the neutrino probes quarks as quasi-free partons, and cross sections are computed using global PDF fits with QCD evolution.  
As the energy decreases toward the GeV scale, however, the partonic description gradually breaks down, and hadronic degrees of freedom reemerge—resonance excitation, meson production, and multinucleon dynamics all contribute significantly.  
This \emph{transition region}, where SIS and DIS overlap, therefore represents one of the most complex and least constrained domains in neutrino–nucleus physics, and remains a critical target for theoretical development and precision measurement in the coming decade.


\subsection{Dominant Nuclear Effects}

When a neutrino scatters from a nucleus, the observed final state reflects not only the underlying weak interaction with individual nucleons but also the complex many-body environment in which those nucleons reside.  
Nuclear effects modify both the magnitude and shape of measurable cross sections, influencing how energy and momentum are distributed among final-state particles.  
They arise from several distinct mechanisms—nuclear binding and Fermi motion, Pauli blocking, meson–exchange currents, nucleon–nucleon correlations, and final-state interactions—each operating on different distance and energy scales.  
Together, these effects determine how the microscopic neutrino–nucleon process is manifested experimentally and represent one of the primary challenges in achieving percent-level precision in oscillation measurements.

\subsubsection{Nuclear Binding and Fermi Motion}

A key feature distinguishing neutrino–nucleus from free-nucleon scattering is that nucleons inside nuclei are both \emph{bound} and \emph{moving}.  
Unlike isolated nucleons at rest, bound nucleons possess intrinsic momenta due to the Pauli exclusion principle and the confining nuclear potential.  
This motion—known as \emph{Fermi motion}—and the associated binding energy play central roles in shaping the kinematics of neutrino–nucleus interactions.  
They broaden energy and momentum transfer distributions, smear sharp kinematic features such as the quasielastic peak, and modify the final-state particle spectra reconstructed in experiments.  

Nucleons inside a nucleus are bound by the strong nuclear force, which lowers their total energy relative to free particles.  
Because of this binding, nucleons are \textit{off their mass shell}, meaning their four-momentum does not satisfy the free-space relation $p^2 = M_N^2$, where $M_N$ is the free nucleon mass.  
Instead, the nucleon’s effective mass in the nuclear medium is modified by the potential $V(r)$ and local density $\rho(r)$, giving rise to an \textit{effective mass} $M^*$ that satisfies $p^2 = M_N^{*2} < M_N^2$. To describe these effects, consider a simplified nuclear Hamiltonian:
\begin{equation}
H = -\frac{\vec{\nabla}^{2}}{2M_N} + V(\vec{r}),
\label{eq:hamiltonian}
\end{equation}
Solving the Schrödinger equation with this Hamiltonian yields the nuclear wave function $\Psi(\vec{r})$, which determines the momentum and energy distributions of the nucleons.  
In the simplest \textit{shell-model} picture, $V(\vec{r})$ is approximated by a smooth, central potential, leading to discrete energy levels occupied by protons and neutrons, as shown in Fig.~\ref{fig:potential_shells} (right).  
More advanced versions include residual interactions that account for nucleon–nucleon correlations, pairing, and collective excitations.  
These refinements are constrained by experimental observables such as binding energies, charge radii, and magnetic moments.  
Within such frameworks, both nuclear binding and Fermi motion naturally emerge from the underlying many-body wave functions.

A more approximate but widely used description—particularly in neutrino event generators—is the Fermi Gas (FG) model.  
Here, nucleons are treated as quasi-free particles confined within a uniform potential well of density $\rho$, with allowed momenta limited by the Fermi momentum:
\[
|\vec{p}\,| \le p_F, \qquad p_F = (3\pi^2 \rho)^{1/3}.
\]
The corresponding momentum distribution is a step function $\Theta(p_F - p)$, and the nucleon energy includes a binding correction.  
In this idealized picture, the nucleus behaves as a degenerate Fermi system: all momentum states up to $p_F$ are occupied, and those above are empty.

A more general and realistic representation is provided by the \emph{spectral function} $S(\vec{p},E)$, which encodes the joint probability of finding a nucleon with momentum $\vec{p}$ and removal energy $E$.  
In the Fermi Gas model this takes the simple form:
\begin{equation}
S(\vec{p},E) \propto 
\Theta(p_F - p)\,
\delta\!\left(E - \sqrt{|\vec{p}|^2 + M^2} + \epsilon\right),
\end{equation}
where $\epsilon$ is the average separation energy of the nucleon. 
In reality, however, the true spectral function is far more intricate, reflecting the shell structure of the nucleus and correlations among nucleons.  
Modern spectral functions are extracted from high-precision electron–nucleus scattering experiments, particularly from exclusive $(e,e'p)$ measurements that directly probe single-nucleon knockout~\cite{JeffersonLabHallA:2022ljj, JeffersonLabHallA:2022cit}.  

\subsubsection{Pauli Blocking}

The Pauli exclusion principle—one of the most fundamental consequences of quantum mechanics—governs the structure of all fermionic systems, including atomic nuclei.  
It states that no two identical fermions can occupy the same quantum state simultaneously.  
In the nuclear context, this principle dictates how protons and neutrons fill discrete energy levels within the nuclear potential well, giving rise to the familiar \textit{shell structure} of nuclei.  
Closed shells correspond to completely filled energy levels, while valence nucleons occupy higher orbitals, influencing key nuclear properties such as spin, parity, and binding energy.

In neutrino–nucleus scattering, the Pauli principle manifests as a suppression of transitions into already-occupied final states—a phenomenon known as \textit{Pauli blocking}.  
When a neutrino interacts with a bound nucleon via exchange of a weak boson ($W^\pm$ or $Z^0$), the struck nucleon can be promoted only to an unoccupied state above the Fermi surface.  
In the Fermi Gas model, all momentum states with $|\vec{p}| \le p_F$ (where $p_F$ is the Fermi momentum) are assumed filled.  
A successful scattering event therefore requires that the final nucleon momentum $p'$ satisfy $p' > p_F$, leaving behind a vacancy—or \textit{hole}—in the nuclear Fermi sea.  
This particle–hole excitation (1p–1h) forms the simplest microscopic picture of quasielastic scattering in nuclei.

From a kinematic perspective, Pauli blocking reduces the available phase space for the outgoing nucleon, particularly at low momentum transfer where few unoccupied states exist above $p_F$.  
This suppression leads to a decrease in the quasielastic cross section at small four-momentum transfer $Q^2$, an effect clearly observed in both electron and neutrino scattering data.  
In the simplest treatments, this can be modeled by multiplying the free-nucleon cross section by a step function enforcing the phase-space constraint:
\[
S(\vec{p}, E)\,\Theta(p' - p_F),
\]
where $S(\vec{p},E)$ is the nucleon spectral function.  
This correction effectively removes contributions from kinematically forbidden transitions into occupied states.

Modern nuclear models refine this simple picture.  
Mean-field and spectral-function approaches incorporate discrete shell structure and realistic energy–momentum distributions, while many-body frameworks include long-range correlations and collective excitations that smooth the sharp Fermi surface.  
In these treatments, Pauli blocking arises naturally from antisymmetrization of the many-body nuclear wave function.  
Despite such refinements, the physical consequence remains the same: at low energy and momentum transfer, available final states are restricted, suppressing scattering rates relative to those on free nucleons. Pauli blocking is therefore a universal feature of neutrino–nucleus interactions, intrinsic to the fermionic nature of nucleons. It is indispensable in any quantitative description of nuclear response functions.

\subsubsection{Meson--Exchange Currents}

\begin{figure}[t]
    \centering
    \includegraphics[width=1.0\textwidth]{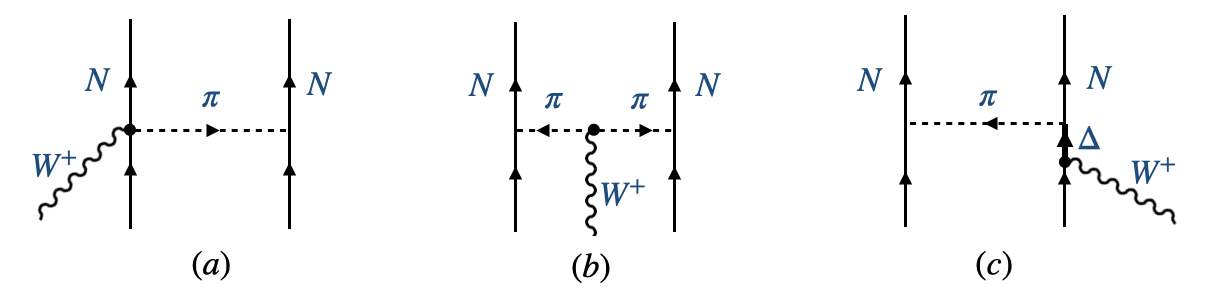}
    \caption{A diagrammatic representation of dominant two-body meson–exchange current (MEC) processes in nuclei: (a) seagull (contact) current, (b) pion–in–flight current, and (c) $\Delta$–isobar current contributions.}
    \label{fig:MEC}
\end{figure}

The nucleons within a nucleus are bound together by the nuclear force—a residual manifestation of the strong interaction that confines quarks within nucleons. At the hadronic scale, this force is effectively mediated by the exchange of virtual light mesons, primarily the $\pi$, $\rho$, and $\omega$, which generate attraction at intermediate distances, as shown in Fig.~\ref{fig:potential_shells} (left). When an external electroweak probe interacts with two correlated nucleons connected by meson exchange, the boson can couple either to the meson itself or to one of the nucleons.  
This generates a two-body current, in contrast to the usual one-body current associated with independent nucleons.  
Such processes can result in the emission of one or more nucleons from the nucleus and are collectively known as \emph{meson–exchange currents} (MECs).  
They represent an essential correction to the simple impulse approximation (IA), in which the probe interacts with only a single nucleon. Such multinucleon processes populate the region between the quasielastic and $\Delta$–resonance peaks and play a crucial role in both interpreting experimental results and refining theoretical models of neutrino interactions.

Depending on the nature of the exchanged meson and the interaction vertex, several characteristic MEC topologies can be identified, the dominant ones are illustrated in Fig.~\ref{fig:MEC}:
\begin{itemize}
    \item \textbf{Seagull (contact) currents}, in which the electroweak boson couples directly to the $\pi NN$ vertex, producing an instantaneous two-nucleon response;
    \item \textbf{Pion–in–flight currents}, where the boson interacts with the virtual pion exchanged between the two nucleons; and
    \item \textbf{$\Delta$–isobar currents}, where the boson excites a nucleon to a virtual $\Delta$ resonance that subsequently decays by exchanging a pion with another nucleon.
\end{itemize}

Additional correlation currents may also arise when one nucleon propagates between the boson–interaction point and the meson–emission vertex. Heavier mesons such as the $\rho$ and $\omega$ can also contribute to MECs, but their effects are generally suppressed because their large masses  
($m_\pi \!\simeq\! 135~\text{MeV}$, $m_\rho \!\simeq\! 775~\text{MeV}$, $m_\omega \!\simeq\! 782~\text{MeV}$) correspond to much shorter interaction ranges.  
At these short distances, the strong repulsive core of the nucleon–nucleon force significantly reduces the associated current amplitudes.  
Consequently, these contributions are often incorporated effectively—for example, through short-range correlation operators or contact terms in many-body calculations.

\subsubsection{Short- and Long-Range Nucleon--Nucleon Correlations}

The nucleons inside a nucleus interact through the strong nuclear force, whose behavior depends sensitively on the distance between them: it is strongly repulsive at short range, attractive at intermediate distances, and negligible at large separations, as shown in Fig.~\ref{fig:potential_shells} (left).  
This intricate balance of forces, combined with the Pauli exclusion principle, gives rise to a relatively large mean free path for nucleons inside the nucleus.  
As a result, the nucleus can often be approximated as a collection of independent particles moving in an average mean-field potential.  
However, correlations among nucleons arise from the complex nature of the nucleon–nucleon ($NN$) interaction.  
These correlations are essential features of nuclear dynamics and are conventionally divided into two broad classes: \textit{long-range correlations (LRCs)} and \textit{short-range correlations (SRCs)}.

\textbf{Long-range correlations} are primarily mediated by pion exchange and extend throughout the nuclear volume.  
They describe collective motion in which many nucleons participate coherently in the nuclear response to an external probe.  
Such correlations allow the energy transferred by an incoming lepton to be redistributed among several nucleons, leading to collective excitations such as Giant Resonances (GR) and low-lying multipole modes (see Fig.~\ref{fig:neutrino_interaction}).  
In theoretical treatments, LRCs are typically incorporated using the Random Phase Approximation (RPA) or related many-body frameworks, which account for coherent superpositions of particle–hole states.  
The inclusion of RPA correlations modifies the nuclear response functions and affects both the shape and magnitude of quasielastic cross sections, often leading to a suppression of the cross section at low momentum transfer.

\textbf{Short-range correlations}, in contrast, arise when two nucleons approach each other closely enough that the repulsive core and tensor components of the $NN$ potential dominate.  
In this regime, the independent-particle picture breaks down: nucleons form strongly correlated pairs with large relative momenta ($|\vec{p}| \gtrsim 300~\text{MeV}/c$) and high removal energies.  
These short-range interactions populate the high-momentum tail of the nuclear momentum distribution and reduce the occupancy of single-particle states relative to mean-field predictions—a reduction quantified by the \emph{spectroscopic factor}.  
High-precision electron-scattering experiments, particularly exclusive $(e,e'p)$ and $(e,e'pp)$ measurements, have provided direct evidence for such correlated nucleon pairs and their high-momentum components, confirming that SRCs are an intrinsic feature of the nuclear ground state rather than small perturbations.

Both SRCs and LRCs play a central role in neutrino–nucleus scattering.  
Long-range correlations influence the collective nuclear response and alter the inclusive cross sections at low energy and momentum transfer, while short-range correlations underpin multinucleon emission channels and strongly affect quasielastic event topologies.  

\subsubsection{Final-State Interactions}

Final-state interactions (FSI) represent one of the most prominent manifestations of nuclear-medium effects in neutrino–nucleus scattering.  
They describe the interactions that outgoing hadrons—nucleons, pions, or other mesons—undergo as they propagate through the residual nucleus after the primary neutrino interaction.  
These secondary interactions can substantially alter the observable final state, changing particle multiplicities, kinematics, and even the reaction topology reconstructed in the detector.

FSI are especially significant in inelastic processes, where mesons produced inside the nucleus may scatter, be absorbed, or undergo charge exchange before escaping.  
Even in quasielastic interactions, FSI can affect the energy and angular distributions of the emitted nucleons.  
For example, a pion created in a resonance decay ($\Delta \!\rightarrow\! N\pi$) may be absorbed by the nuclear medium through processes such as $\pi NN \!\rightarrow\! NN$, leaving only nucleons in the final state.  
Experimentally, such events are indistinguishable from genuine QE interactions and thus contribute to the so-called QE-like event sample.  
This ambiguity complicates the interpretation of measured cross sections and challenges the separation of reaction mechanisms. When the final hadronic state consists only of nucleons, FSI are typically modeled by propagating the outgoing nucleon(s) through an effective nucleon–nucleon potential that incorporates both mean-field and correlation effects.  
This treatment accounts for rescattering, deflection, energy loss, and absorption as the ejected nucleons traverse the nuclear medium. 

In pion-production channels, analogous FSI processes include:
\begin{itemize}
    \item \textbf{Pion absorption:} $\pi NN \!\rightarrow\! NN$, which removes the pion from the final state;
    \item \textbf{Elastic scattering:} $\pi N \!\rightarrow\! \pi N$, which alters pion kinematics without changing its identity;
    \item \textbf{Charge exchange:} $\pi^- p \!\rightarrow\! \pi^0 n$ (and analogs), which change the pion’s charge and thereby the final-state topology.
\end{itemize}
Among these, pion absorption has particularly important experimental consequences.  
It can artificially enhance the apparent QE signal—since non-QE events are misclassified as QE-like—and distort the reconstructed neutrino energy when QE kinematics are assumed.  

Modern theoretical treatments incorporate FSI through several complementary approaches. Intranuclear cascade models simulate hadron propagation and reinteractions using semiclassical transport equations. Optical potential approaches describe FSI via complex mean-field potentials encoding both absorption and elastic scattering. Each method offers a different balance between computational efficiency and physical completeness, and comparisons with benchmark data from electron scattering and pion–nucleus experiments serve as essential validation tests.

Together, the nuclear effects discussed above -- binding, Fermi motion, Pauli blocking, meson–exchange currents, nucleon–nucleon correlations, and final-state interactions -- define the rich nuclear environment in which neutrino interactions occur. A quantitative description of these processes requires theoretical frameworks that consistently incorporate both nuclear dynamics and electroweak interaction physics. The following section surveys the principal theoretical approaches developed to model these effects across medium energies.


\section{Theory Landscape: Foundational and Widely Used Approaches}\label{sec:theory}

The theoretical description of neutrino–nucleus interactions remains one of the most demanding problems in modern nuclear and particle physics.  
Modeling neutrino–nucleus scattering in the medium-energy regime is particularly challenging because of the stark contrast between the interacting systems.  
The incoming lepton can be treated as a relativistic point-like spin-$\tfrac{1}{2}$ particle, whereas the target nucleus is a correlated quantum many-body system exhibiting shell structure, pairing, and collective excitations.  
To make progress, theorists employ a hierarchy of approximations.  
The relevant degrees of freedom—whether nucleons, mesons, or quarks—depend strongly on the energy and momentum transfer, and any comprehensive description must transition smoothly among these regimes.
As a result, realistic calculations must rely on effective theories or phenomenological models that bridge QCD with observable nuclear phenomena.  

Over the last two decades, the field has made rapid progress.  
A comprehensive suite of models now exists to study neutrino–nucleus cross sections across a broad range of energies.  No single model provides a universal description across all energy domains; rather, each captures key physical mechanisms within its region of validity~\cite{Ruso:2022qes}. Comparisons among these models—and with high-precision experimental data—have provided stringent tests of nuclear theory and guided the development of improved event generators.  
Any realistic description of $\nu A$ interactions must therefore include a consistent treatment of long- and short-range correlations, meson–exchange currents, and final-state interactions, each of which modifies both the magnitude and shape of observable cross sections.  
The sections that follow survey these foundational approaches, highlighting their physical assumptions, domain of applicability, and relevance for ongoing and future neutrino experiments.

\subsection{Fermi Gas Based Approaches}\label{sec:LDA}

The Fermi Gas (FG) based models built on the Local Density Approximation (LDA) provides one of the most practical and widely used frameworks for describing neutrino–nucleus interactions in the intermediate-energy regime.  
In this approach, the nucleus is treated as locally uniform nuclear matter characterized by position-dependent proton and neutron densities, $\rho_{p,n}(r)$, derived from experimental charge distributions.  
Rather than resolving discrete shell effects, the LDA averages over the local environment, enabling efficient and physically motivated modeling of inclusive observables.  
Although it lacks the fine-grained shell structure required for describing exclusive reactions, its balance of realism and computational simplicity has made it the foundation of most modern neutrino–nucleus event generators.

The FG models treat nucleons as independent fermions moving freely in a constant potential well up to a local Fermi momentum $p_F(r)$.  
In the global FG model, both the nuclear density and $p_F$ are constant, whereas in the local version they vary with the nuclear radius, reflecting the spatial dependence of $\rho(r)$.  
The ground-state momentum distribution is described by a step function $\Theta(p_F - p)$, while knocked-out nucleons occupy states above the Fermi surface consistent with Pauli blocking.  
The Fermi momentum is directly related to the local nuclear density through
\begin{equation}
p_F(r) = \hbar \left[3\pi^2 \rho(r)\right]^{1/3},
\label{eq:pF}
\end{equation}
where $\rho(r)$ is the nuclear density.  
The nucleon momentum distribution in the ground state is then approximated by
\begin{equation}
n(\vec{p},r) = \Theta\!~\big(p_F(r) - |\vec{p}|\big),
\end{equation}
so that all states with $|\vec{p}| < p_F(r)$ are occupied, while those above are empty.  
This defines a locally degenerate Fermi sea, with Pauli blocking enforced by the exclusion of transitions into already occupied states.  
Although the FG framework neglects binding-energy variations and shell effects, it remains the starting point for most neutrino event generators owing to its analytic tractability and straightforward implementation. These simulations support a two-step picture: a primary neutrino–nucleon interaction at position $r$, followed by an intranuclear cascade describing the propagation and reinteraction of outgoing hadrons through the residual medium.

Several key implementations of the LDA have extended its predictive power.  
The Valencia model~\cite{Nieves:2004wx, Nieves:2011pp, Nieves:2011yp, Nieves:2013fr} and the Lyon model~\cite{Martini:2009uj, Martini:2011wp, Martini:2013sha, Martini:2014dqa, Martini:2016eec} build on the local Fermi gas framework but incorporates collective nuclear excitations via the Random Phase Approximation (RPA).  
In these approaches, long-range correlations between particle–hole and $\Delta$–hole excitations redistribute nuclear response strength, especially at low momentum transfer.  
Medium modifications are introduced through complex self-energies, which account for collisional broadening and changes in the nucleon dispersion relation.  

\subsection{Scaling and Superscaling Approaches}
\label{sec:SuSA}

Scaling phenomena signify the emergence of universal patterns in observables that become independent of certain kinematic parameters.  
In lepton–nucleus scattering, scaling arises when, for sufficiently large momentum transfers ($|\vec{q}| \gtrsim 400$–$500~\text{MeV}/c$), the quasielastic cross section factorizes into a single-nucleon term multiplied by a universal nuclear function.  
This function, the \textit{scaling function}, encapsulates the essential nuclear dynamics and depends on a dimensionless scaling variable $\psi$, defined in terms of $(\omega,\vec{q})$.  
When $f(\psi)$ becomes independent of $|\vec{q}|$, the system exhibits \textit{scaling of the first kind}; when the same function describes different nuclei, it exhibits \textit{scaling of the second kind}.  
The simultaneous realization of both defines \textit{superscaling} (SuSA)~\cite{Donnelly:1998xg, Donnelly:1999sw}. In practice, the measured inclusive cross section is divided by the corresponding single-nucleon cross section to extract a reduced, or ``scaling,'' function, which, when plotted versus $\psi$, demonstrates scaling of both the first and second kinds.  
Their simultaneous manifestation constitutes the observed \textit{superscaling behavior}.  

The relativistic Fermi gas (RFG) model provides a natural theoretical setting in which both scaling and superscaling occur exactly.  
Despite its simplicity, it captures the main features of QE scattering and yields analytic expressions for inclusive cross sections and nuclear response functions.  
Within the RFG, the dimensionless scaling function is given by  
\begin{equation}
f(\psi) = k_F \,
\frac{d^2\sigma / (d\Omega\, d\omega)}
{\sigma_{\text{Mott}}\,(v_L G_L + v_T G_T)} ,
\label{eq:superscaling}
\end{equation}
where $k_F$ is the Fermi momentum, $\sigma_{\text{Mott}}$ is the Mott cross section, and $v_L$ and $v_T$ are kinematic factors associated with the longitudinal and transverse response channels.  
The quantities $G_L$ and $G_T$ denote the corresponding single-nucleon response functions, evaluated consistently within the RFG formalism.  

Analyses of extensive inclusive $(e,e')$ data—particularly on $^{12}$C—have confirmed that while the RFG reproduces the qualitative scaling trends, it fails to describe quantitatively the magnitude and shape of the measured longitudinal scaling function.  At energy transfers above the QE peak, violations of scaling arise from processes beyond the impulse approximation, such as meson–exchange currents, two-body correlations, and inelastic excitations.  
The SuSA framework has therefore been extended to the $\Delta$-resonance and inelastic regions.  
A refined version, \textit{SuSAv2}, incorporates relativistic mean-field (RMF) dynamics within the superscaling formalism, including the effects of scalar and vector potentials on bound and outgoing nucleons while maintaining current conservation~\cite{Gonzalez-Jimenez:2014eqa, Megias:2016fjk}. These models provide a reliable and computationally efficient representation of inclusive lepton–nucleus scattering.  
Their predictive power and simplicity have led to widespread implementation in modern neutrino event generators, where they serve as essential benchmarks for testing the consistency of microscopic nuclear models in neutrino oscillation analyses.  

\subsection{Mean-Field Approaches}
\label{sec:MF}

\begin{figure}[t]
    \centering
    \includegraphics[width=1.0\textwidth]{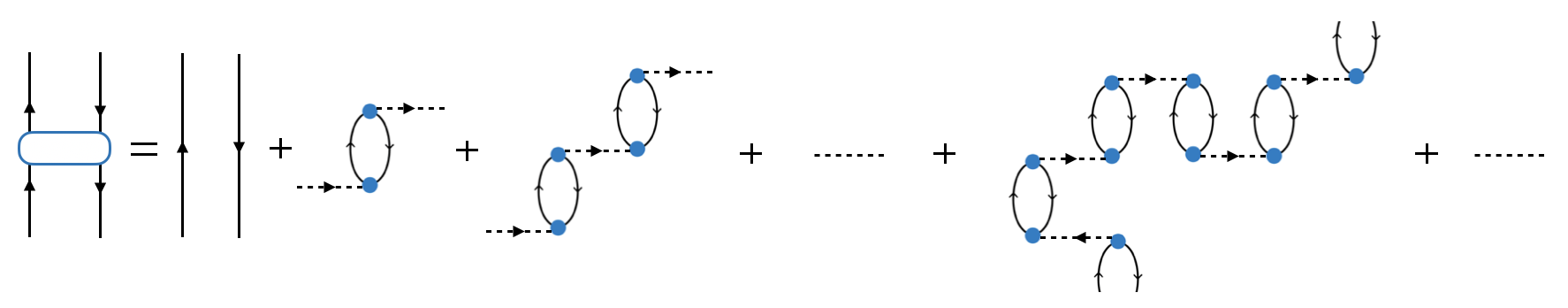}
    \caption{A diagrammatic representation of the random-phase approximation (RPA) as a sum of ring diagrams.}
    \label{fig:RPA_diagrams}
\end{figure}

As discussed earlier, the many-body Hamiltonian describing a nucleus of $A$ interacting nucleons can be written as
\begin{equation}
H = \sum_i H^{[1]}(i) + \sum_{i \neq j} v(i,j),
\label{eq:manybody_H}
\end{equation}
where $H^{[1]}(i)$ is the single-particle Hamiltonian and $v(i,j)$ the two-body interaction potential.  
A widely used approach to this problem is the \textit{Hartree--Fock (HF) approximation}, which is an important way of proceeding beyond perturbation theory that occurs in the quantum theory of many-body systems.  
This method starts by introducing a convenient decomposition of the Hamiltonian:
\begin{align}
H &= \sum_i T(i) + \sum_{i \neq j} V(i,j), \label{eq:H-decompose}
\end{align}
where $T(i)$ is the single-particle kinetic energy operator and $V(i,j)$ is some approximation to the NN potential. One adds and subtracts some single particle potential $U(i)$ representing an average or \textit{mean-field} potential, thereby defining the single particle hamiltonian as
\begin{align}
H^{[1]}(i) &\equiv T(i) + U(i), \label{eq:single_H}\\
v(i,j) &\equiv V(i,j) - U(i), \label{eq:residual_V}.
\end{align}

In the HF method, one begins with a trial wave function represented as a single Slater determinant:
\begin{equation}
|F\rangle = |\Phi_0\rangle = |\alpha_1 \alpha_2 \ldots \alpha_A\rangle,
\label{eq:Slater_det}
\end{equation}
where the lowest single-particle levels $\alpha_i$ are filled up to the Fermi level, as shown in Fig.~\ref{fig:potential_shells} (right).  
Particle–hole excitations above this Fermi sea can be represented as
\begin{equation}
|ph\rangle = a^\dagger_{\alpha_p} b^\dagger_{\alpha_h} |\Phi_0\rangle,
\label{eq:ph_state}
\end{equation}
where $a^\dagger_{\alpha_p}$ and $b^\dagger_{\alpha_h}$ are particle and hole creation operators, respectively.  
These $1p$--$1h$ excitations form the building blocks for more complex configurations such as $2p$--$2h$, $3p$--$3h$, etc.

In essence, the HF method provides an optimized mean-field (MF) description where each nucleon moves in an average potential generated by all others, and the residual interaction accounts for correlations beyond the independent-particle approximation. 
This consistent treatment of the initial and final states yields a quantum-mechanical description of 
Pauli blocking and elastic final-state interactions, capturing nucleon distortion as the 
ejected particle propagates through the nuclear medium.

Constructing a realistic nuclear potential remains a central challenge.  
Phenomenological models employ parameterized Woods–Saxon or harmonic-oscillator potentials fitted to nuclear properties, whereas microscopic schemes derive $U$ self-consistently from effective nucleon–nucleon interactions.  
Skyrme-type forces are particularly common, adjusted to reproduce binding energies, charge radii, and single-particle spectra.  
The resulting HF potentials describe both bound and continuum states consistently and form the basis of several modern mean-field frameworks, such as the \textit{Ghent model}, which employs Skyrme-based wave functions for initial and final nucleon state. The same potential is employed for both bound and
continuum states, ensuring orthogonality of wave functions and consistency in the description of initial and final statess~\cite{Jachowicz:2002rr, Pandey:2014tza, Pandey:2016jju, VanCuyck:2016fab, Nikolakopoulos:2019qcr}. 

The \textit{Random-Phase Approximation} (RPA) extends the HF framework to include small-amplitude collective excitations.  
While HF captures the average nuclear potential, RPA incorporates residual correlations between particle–hole ($p$–$h$) excitations.  
Diagrammatically, it corresponds to the infinite summation of ring diagrams, shown schematically in Fig.~\ref{fig:RPA_diagrams}, each ring representing repeated $p$–$h$ interactions mediated by the residual force. The upward lines denote particles (states above the Fermi surface), while the downward lines
denote holes (states below the Fermi surface). The name “random-phase” arises because, under certain assumptions,
off-diagonal matrix elements of the interaction acquire random phases and can be averaged statistically, simplifying the formalism while
preserving key physical correlations.
A continuum version, the \textit{CRPA}, extends RPA to include coupling to unbound states using the same Skyrme interaction as the HF ground state. This resummation captures collective modes such as giant resonances and modifies the nuclear response functions.   
The HF–CRPA framework successfully describes processes from low-energy excitations, giant resonances, to a significant part of the medium energy spectrum.  

In addition, the \textit{Relativistic Mean Field} (RMF) approaches use complex optical potentials while restoring flux conservation through an explicit summation over all final states ~\cite{Gonzalez-Jimenez:2019qhq, Gonzalez-Jimenez:2019ejf, Franco-Munoz:2023zoa}.  
Although MF approaches break the simple factorization structure of the cross section and are more computationally demanding than Fermi gas based models, MF approaches provide a fully quantum-mechanical treatment of initial- and final-state interactions and have achieved broad success across kinematic regimes.  
Frameworks such as HF–CRPA and RMF therefore play a pivotal role in bridging microscopic nuclear dynamics with experimental observables in current and next-generation neutrino experiments.

\subsection{Spectral Function Approach}
\label{sec:SF}

The spectral function (SF) encodes the probability of removing a nucleon with momentum $\vec{k}$ and separation energy $E$, 
leaving the residual nucleus with excitation energy $E_{\mathcal{R}} = E_0 - m + E$. The SF formalism provides a powerful and versatile framework for describing lepton–nucleus interactions, where the nuclear final state can be 
factorized into a lepton–nucleon vertex and a residual $(A-1)$ system.  

In the shell-model framework—where nucleons move independently in a mean-field potential—the
spectral function takes the simplified form 
\begin{equation}
S_{\mathrm{SM}}(\vec{k},E) =
\sum_{n \in \{F\}} 
|\phi_n(\vec{k})|^2 \, \delta(E - E_n),
\label{eq:ssm}
\end{equation}
where $\phi_n(\vec{k})$ is the momentum-space wave function of occupied single-particle levels 
and $E_n$ their corresponding energy eigenvalues~\cite{Benhar:2006wy}. Beyond the mean-field picture, the hole spectral function $S_h(E,\vec{k})$ generalizes this concept as  
\begin{equation}
S_h(\vec{k},E) = 
\sum_n 
\big| 
\langle R_{A-1}(-\vec{k}) | a_{\vec{k}} | I_A \rangle 
\big|^2 
\delta(E - E_n + E_I)
= \frac{1}{\pi}\, \mathrm{Im}\, G_h(\vec{k},E),
\label{eq:SF_def}
\end{equation}
where $G_h(\vec{k},E)$ is the hole Green’s function describing the propagation of the removed nucleon.  
The SF naturally separates into mean-field and correlation components: the MF term encodes the shell structure of the nucleus,
with nucleons occupying discrete orbitals constrained by the Pauli principle, while the correlation term 
accounts for short-range and tensor correlations responsible for the high-momentum tail ($|\vec{k}| > k_F$).  

A widely used implementation is the Benhar \textit{et al.} model~\cite{Benhar:2005dj}, which combines empirical $(e,e'p)$ inputs with 
theoretical correlation calculations.  
The MF part is constrained by $(e,e'p)$ data and employs shell-model wave functions with quenched spectroscopic factors and energy broadening, 
while the correlation contribution is derived from Correlated Basis Function (CBF) theory in nuclear matter 
and extended to finite nuclei via a local-density approximation (LDA). In practical applications, the SF is typically implemented within the plane-wave impulse approximation (PWIA), treating the ejected nucleon as a free particle.  
Final-state interactions (FSIs) are incorporated by shifting the nucleon energy using a real optical potential 
$U$, or by applying a folding function derived from Glauber theory to account for multiple scattering.  
These corrections reproduce the broadening and quenching of the QE peak observed in inclusive $(e,e')$ spectra~\cite{Benhar:2005dj, Ankowski:2014yfa, Rocco:2015cil}. Recent advances in nuclear many-body theory have further refined the SF approach.  
Quantum Monte Carlo (QMC) methods now compute $S_h(\vec{k},E)$ for light nuclei ($A \leq 12$) using 
variational Monte Carlo overlaps and two-nucleon momentum distributions to isolate short-range correlations~\cite{Barbieri:2019ual, Rocco:2019gfb}.

\subsection{Ab-Initio Approaches}
\label{sec:abinitio}

Recent advances in high-performance computing and theoretical nuclear physics have enabled a new 
generation of \textit{ab initio} (“from first principles”) approaches that solve the nuclear many-body 
problem in a systematically improvable and quantitatively controlled way.  
These methods aim to describe nuclear structure and dynamics directly from interactions among individual 
nucleons—protons and neutrons—treated as the fundamental degrees of freedom.  
The nucleus is modeled as a system of $A$ nonrelativistic nucleons governed by the Hamiltonian  
\begin{equation}
H = \sum_{i=1}^{A} \frac{\mathbf{p}_i^2}{2m} 
    + \sum_{i<j} V_{ij} 
    + \sum_{i<j<k} V_{ijk},
\label{eq:abinitio_hamiltonian}
\end{equation}
where $V_{ij}$ and $V_{ijk}$ denote two- and three-nucleon (3N) interactions, respectively.  

Early \textit{ab initio} studies employed phenomenological nucleon–nucleon potentials such as 
Argonne~v18 (AV18)~\cite{Wiringa:1994wb}, which combines long-range one-pion exchange with 
short-range components representing multi-meson exchanges.  
More recently, the advent of chiral Effective Field Theory ($\chi$EFT)~\cite{Weinberg:1990rz,Epelbaum:2008ga} 
has provided a systematic expansion of nuclear forces and currents consistent with Quantum Chromodynamics (QCD).  
In $\chi$EFT, the nuclear Hamiltonian and electroweak operators are expanded in powers of 
$Q/\Lambda_\chi$, where $Q$ is the typical momentum scale and $\Lambda_\chi \sim 1$~GeV the chiral-symmetry-breaking scale:
\begin{equation}
H = H_{\text{LO}} + H_{\text{NLO}} + H_{\text{N}^2\text{LO}} + \cdots ,
\label{eq:chiral_expansion}
\end{equation}
yielding a hierarchy of one-, two-, and three-body contributions that can be systematically improved and 
quantified in uncertainty.  
This framework establishes a direct link between nuclear dynamics and the underlying symmetries of QCD.

Several complementary \textit{ab initio} techniques have been developed to compute nuclear 
electroweak response functions.  
Their common strength lies in the consistent treatment of nuclear wave functions and current operators, 
ensuring that one- and two-body correlations are included on equal footing. The Green’s Function Monte Carlo (GFMC) method provides essentially exact solutions for light nuclei 
($A \lesssim 12$) with realistic NN and 3N forces~\cite{Lovato:2016gkq,Lovato:2020kba}. The Coupled-Cluster (CC) approach extends the reach to medium-mass nuclei ($A \sim 40$) by expressing 
the correlated wave function as an exponential ansatz,
\begin{equation}
|\Psi\rangle = e^{T} |\Phi_0\rangle, \qquad 
T = T_1 + T_2 + \cdots ,
\label{eq:cc_ansatz}
\end{equation}
where $T_n$ represents $n$-particle–$n$-hole excitation operators acting on a reference Slater determinant $|\Phi_0\rangle$.  
Truncations at finite excitation rank yield a hierarchy of approximations~\cite{Hagen:2013nca,Sobczyk:2021dwm,Acharya:2024xah}. The Short-Time Approximation (STA)~\cite{Pastore:2019urn} extends \textit{ab initio} calculations 
to higher momentum transfers by factorizing short-time two-body dynamics from long-range propagation, 
retaining key correlation and MEC physics while substantially reducing computational cost.  

A defining feature of modern \textit{ab initio} approaches is their ability to quantify theoretical 
uncertainties, including those arising from truncations in the $\chi$EFT expansion and from low-energy constants.  
These uncertainties are now propagated to observables, enabling statistically meaningful comparisons 
with experimental data.  
The combination of realistic nuclear forces, consistent electroweak operators, and quantified uncertainties 
makes \textit{ab initio} methods an indispensable benchmark for neutrino–nucleus interaction modeling.


\section{Experimental Landscape: Current and Upcoming Experiments}\label{sec:exp}

Neutrino–nucleus scattering has been explored for more than half a century across both charged-current and neutral-current channels, and using a wide range of nuclear targets—from hydrogen and deuterium to heavier nuclei such as carbon, oxygen, iron, and argon.  
Early measurements, performed with bubble chambers and spark counters, provided the first systematic evidence for weak interactions and established the qualitative features of quasielastic, resonance, and deep-inelastic processes.  
However, these pioneering experiments suffered from limited statistics, uncertain flux normalizations, and incomplete kinematic coverage, leaving significant ambiguities in absolute cross sections and in the detailed understanding of nuclear effects.

The modern era of neutrino physics has been defined by the advent of high-intensity, accelerator-based neutrino beams optimized for oscillation studies.  
These facilities—capable of delivering well-characterized fluxes with energies from hundreds of~MeV to several~GeV—have enabled precise differential measurements of neutrino–nucleus cross sections on a variety of targets.  
In contrast to early inclusive measurements, contemporary experiments employ fine-grained detectors capable of reconstructing both leptonic and hadronic final states, allowing direct study of the nuclear mechanisms that shape observable distributions.  

The experimental exploration of neutrino oscillations now rests on a global network of facilities spanning multiple energy regimes and baselines.  
These programs collectively probe both \emph{appearance} and \emph{disappearance} channels and rely on comparisons between near and far detectors.  
However, because all detectors employ nuclear targets, the achievable precision in oscillation parameters is fundamentally limited by uncertainties in neutrino–nucleus interactions.  
A major challenge is the mapping between the \textit{true} and \textit{reconstructed} neutrino energy: nuclear effects redistribute the visible energy, often biasing the reconstructed spectra and thereby the inferred oscillation parameters.

Modern near detectors now function as precision scattering facilities, providing differential cross-section data on various nuclei to benchmark theoretical models and constrain flux and detector effects.  
Subtle differences between neutrino and antineutrino interactions, as well as between $\nu_\mu$ and $\nu_e$ channels, further complicate interpretation.  
Extrapolating measurements between detectors demands a detailed understanding of how cross sections vary with both energy and target composition.

In summary, the present experimental landscape combines long-baseline oscillation programs, dedicated cross-section experiments, and complementary electron–nucleus studies.  
The following subsections summarize major short- and long-baseline facilities, specialized cross-section measurements, and ongoing efforts to benchmark neutrino data with precision electron-scattering results.

\begin{table}[ht]
\centering
\caption{Summary of current and upcoming medium-energy neutrino experiments, organized by beam energy, target material, and detector technology.}
\begin{tabular}{| l | c | c | c | c | c |}
\hline
\textbf{Experiment} & \textbf{Flavor} & \textbf{$\nu_\mu$ Flux Peak (GeV)} & \textbf{Target(s)} & \textbf{Detection} & \textbf{Run Period} \\
\hline\hline
\multicolumn{6}{|l|}{\textbf{Short-Baseline Experiments}}\\
\hline
MicroBooNE & $\nu_\mu,\nu_e$ & 0.8, 0.3 & Ar & Tracking + Calorimetry & 2015--2020 \\
SBND       & $\nu_\mu,\nu_e$ & 0.8 (PRISM: 0.6--0.8) & Ar & Tracking + Calorimetry & 2024-- \\
ICARUS     & $\nu_\mu,\nu_e$ & 0.8, 0.3 & Ar & Tracking + Calorimetry & 2022-- \\
\hline\hline
\multicolumn{6}{|l|}{\textbf{Long-Baseline Experiments}}\\
\hline
T2K   & $\nu_\mu,\bar\nu_\mu,\nu_e,\bar\nu_e$ & 0.6, 0.8, 1.0 & CH, H$_2$O, Fe & Tracking & 2010-- \\
NOvA  & $\nu_\mu,\bar\nu_\mu,\nu_e,\bar\nu_e$ & 2.0 & CH$_2$ & Tracking + Calorimetry & 2010-- \\
T2HK  & $\nu_\mu,\bar\nu_\mu,\nu_e,\bar\nu_e$ & 0.6, 0.8, 1.0 & CH, H$_2$O, Fe & Tracking & Upcoming \\
DUNE  & $\nu_\mu,\bar\nu_\mu,\nu_e,\bar\nu_e$ & 2.5 (PRISM: 0.5--2.5) & H, C, Ar & Tracking + Calorimetry & Upcoming \\
\hline\hline
\multicolumn{6}{|l|}{\textbf{Dedicated Cross-Section Experiments}}\\
\hline
MINERvA & $\nu_\mu,\bar\nu_\mu,\nu_e,\bar\nu_e$ & 3.5, 6.0 & He, C, CH, H$_2$O, Fe, Pb & Tracking + Calorimetry & 2009--2019 \\
ANNIE   & $\nu_\mu$ & 0.8 & CH, H$_2$O & Cherenkov & 2019-- \\
NINJA   & $\nu_\mu,\bar\nu_\mu,\nu_e,\bar\nu_e$ & 1.0 & CH, H$_2$O, Fe & Emulsion & 2015-- \\
\hline
\end{tabular}
\label{tab:expt}
\end{table}

\subsection{Short-Baseline Experiments}\label{subsec:sbn}

The Short-Baseline Neutrino (SBN) program at Fermilab represents a cornerstone of modern experimental neutrino physics.  
It comprises three liquid-argon time projection chambers (LArTPCs)—SBND (near), MicroBooNE (intermediate), and ICARUS (far)—aligned along the Booster Neutrino Beam (BNB)~\cite{MicroBooNE:2015bmn}.  
Together, these detectors span baselines from 110~m to 600~m, providing a well-controlled environment for both precision cross-section measurements and sensitive searches for sterile neutrinos at the $\mathcal{O}(\text{eV})$ scale.  
MicroBooNE and ICARUS also sample the off-axis NuMI flux, extending their energy coverage and enabling complementary analyses.  
Beyond oscillation searches, SBN delivers a comprehensive program of $\nu$–Ar scattering measurements directly relevant to DUNE, with its high-resolution LArTPC imaging capability allowing detailed studies of quasielastic, multinucleon, resonance, and shallow inelastic interactions.

\subsubsection{MicroBooNE}
Located 470~m downstream of the BNB target, MicroBooNE was the first operational LArTPC in the U.S., featuring an 85~ton active volume and a total exposure of $12.5\times10^{20}$~protons-on-target (POT), corresponding to roughly half a million neutrino interactions.  
In addition to the on-axis BNB flux, it recorded off-axis NuMI events, extending its kinematic reach to higher energies.  
MicroBooNE has produced an extensive portfolio of inclusive and exclusive $\nu$–Ar cross-section measurements, spanning final states with various proton multiplicities~\cite{MicroBooNE:2023tzj, MicroBooNE:2022emb, MicroBooNE:2020akw}, pion-production channels~\cite{MicroBooNE:2018neo}, and rare processes such as $\eta$ and hyperon ($\Lambda$) production~\cite{MicroBooNE:2022cls, MicroBooNE:2023ubu}.  
An example of a reconstructed electron-neutrino charged-current single-pion interaction is shown in Fig.~\ref{fig:neutrino_events} (right), where an electromagnetic shower—consistent with an outgoing electron—is accompanied by a charged pion track.  
MicroBooNE has also pioneered precision reconstruction techniques that set the stage for future LArTPC experiments.

\subsubsection{SBND}
The SBND detector, positioned only 110~m from the BNB target, contains a 112~ton active LArTPC volume and records approximately $7\times10^3$ events per day—about $2\times10^6$ $\nu_\mu$ and $1.5\times10^4$ $\nu_e$ per year.  
Prior to DUNE, SBND will deliver the most precise inclusive and exclusive $\nu$–Ar cross sections, with fine spatial granularity enabling short-track proton reconstruction (relevant for 2p–2h and FSI studies) and robust $e/\gamma$ separation for $\nu_e$ samples.  
The fine spatial resolution of the LArTPC technology allows detailed reconstruction of final-state topologies, enabling SBND to probe subtle nuclear effects and rare processes such as hyperon ($\Lambda$, $\Sigma^+$) production in neutrino–argon scattering~\cite{MicroBooNE:2015bmn, SBND:2025lha}.  
A PRISM-like capability is realized by analyzing interactions across effective off-axis angles ($\sim0.2^\circ$–$1.6^\circ$), sampling spectra that differ in peak energy and high-energy tails by roughly 200~MeV~\cite{SBND:2025plf}.  
SBND began physics data-taking in late~2024 and is expected to record approximately ten million interactions in its lifetime.

\subsubsection{ICARUS}
ICARUS, originally constructed at Gran Sasso and later refurbished at CERN, was installed at Fermilab in 2022 as the far detector of the SBN program.  
It samples both the on-axis BNB and a $\sim$103~mrad off-axis NuMI flux, the latter providing a valuable $\nu_e$-enriched component arising from kaon and muon decays.  
ICARUS is designed to measure inclusive and exclusive $\nu_e$ and $\nu_\mu$ interactions across QE-like, RES, and DIS regimes.  
With its large 470~ton active volume and high imaging resolution, ICARUS bridges the energy range between the SBN program and the low-energy frontier of DUNE, providing essential inputs for cross-section modeling and oscillation analyses~\cite{ArteroPons:2025atg}.

\begin{figure}[t]
    \centering
    \includegraphics[width=0.43\textwidth]{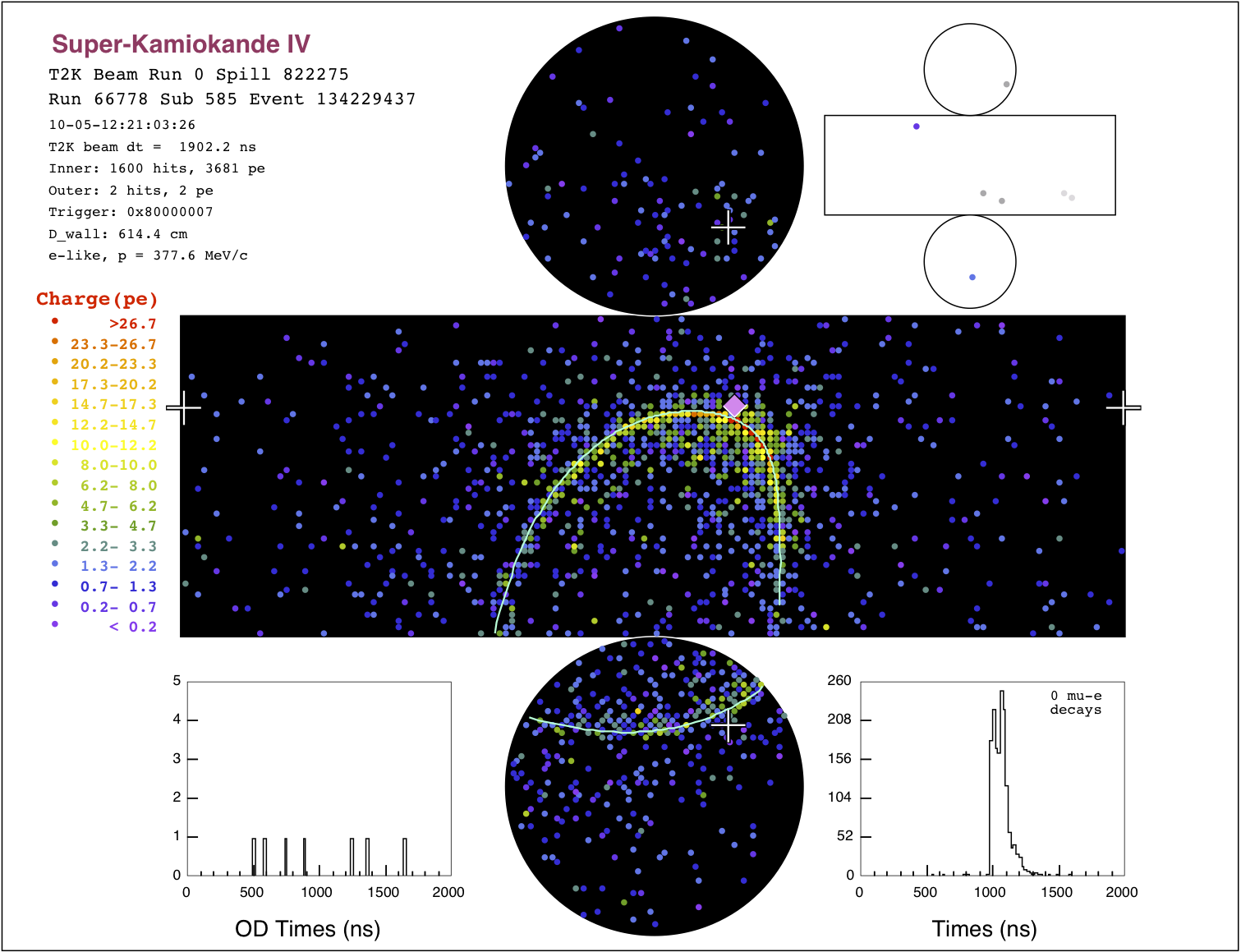}
    \includegraphics[width=0.50\textwidth]{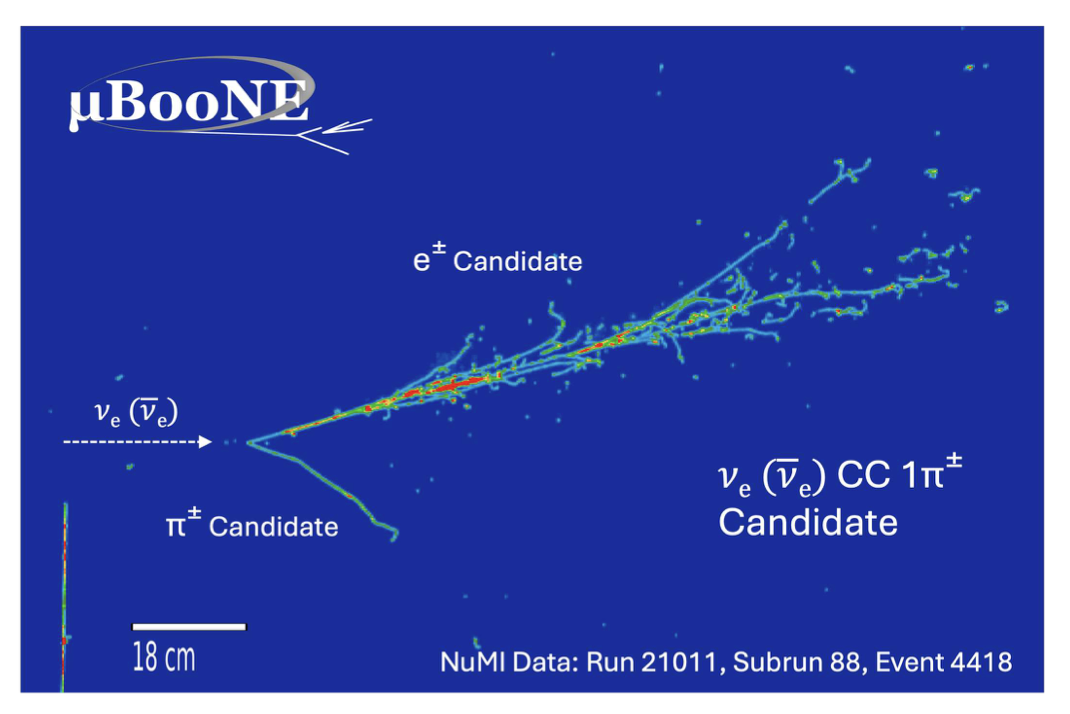}
    \caption{Examples of $\nu_e$–nucleus scattering events as observed in (left) the Super-Kamiokande water Cherenkov detector and (right) the MicroBooNE liquid argon time projection chamber. Images from Refs.~\cite{T2K:SuperK_event_displays, MicroBooNE:2025pvb}.}
    \label{fig:neutrino_events}
\end{figure}

\subsection{Long-Baseline Experiments}\label{subsec:lbl}

Long-baseline neutrino experiments are designed to measure oscillation parameters—particularly leptonic CP violation, mass ordering, and mixing angles—by comparing the flavor composition of neutrino beams at near and far detectors separated by hundreds of kilometers.  
Because all detectors use complex nuclear targets, a deep understanding of neutrino–nucleus interactions across different energy regimes is essential to interpret observed spectra and control systematic uncertainties.  
The following overview summarizes the key features and cross-section programs of the T2K/Hyper-Kamiokande, NOvA, and DUNE experiments.

\subsubsection{T2K}
The Tokai-to-Kamioka (T2K) experiment uses a 295~km baseline from the J-PARC accelerator to the Super-Kamiokande (SK) detector.  
Its near-detector suite samples complementary neutrino spectra at different off-axis angles.  
ND280, located at $2.5^\circ$ (peaked near $E_\nu\!\sim\!0.6$~GeV), provides detailed measurements of CC0$\pi$ (QE-like), CC1$\pi$, and inclusive cross sections on hydrocarbon, with analyses that leverage proton and pion kinematics to probe nuclear effects and final-state interactions (FSI).  
The on-axis INGRID detector monitors the flux, while WAGASCI + BabyMIND at $1^\circ$ compares water and scintillator targets with magnetized charge-sign separation.  
The ND280 upgrade adds low-threshold proton/pion tracking, broader angular coverage, and neutron time-of-flight capability, improving sensitivity to 2p–2h and $\nu_e/\bar{\nu}_e$ cross sections and enabling joint fits across detectors to reduce flux–cross-section degeneracies~\cite{T2K:2018rnz, T2K:2018lnf, T2K:2019ddy, T2K:2020jav}.  
Figure~\ref{fig:neutrino_events}~(left) shows an example of an electron-like neutrino interaction event in the Super-Kamiokande water Cherenkov detector. 

\subsubsection{NOvA}\label{subsec:nova-nd}
The NOvA experiment employs the off-axis NuMI beam at Fermilab, delivering a narrow energy spectrum peaked near $E_\nu \!\sim\! 2$~GeV to a 14~kt far detector located 810~km away.  
The near and far detectors are segmented liquid-scintillator calorimeters with identical designs, enabling precise near-to-far extrapolation with minimal model dependence.  
The near detector provides high-statistics, finely binned cross sections (1--3~GeV) across QE, MEC, and RES regimes, with dedicated $\nu$/$\bar\nu$ running for inclusive and exclusive $\nu_e$CC, $\nu_\mu$CC, and NC channels~\cite{NOvA:2019bdw, NOvA:2021eqi, NOvA:2022see}.  
Its extensive carbon-target data serve as a key reference for model development.

\subsubsection{T2HK}
Hyper-Kamiokande (HK) builds directly on T2K and SK experience to deliver precision measurements of CP violation and the neutrino mass ordering, using both accelerator and atmospheric neutrinos.  
Its CP sensitivity relies on accurate modeling of $\nu_\mu\!\rightarrow\!\nu_e$ and $\bar{\nu}_\mu\!\rightarrow\!\bar{\nu}_e$ appearance, particularly $\nu_e/\bar{\nu}_e$ cross-section systematics and single-pion production at low energies.  
HK will employ two complementary near detectors: the upgraded ND280 and the Intermediate Water Cherenkov Detector (IWCD).  
The IWCD, located $\sim$1~km from the target, can move vertically to sample different off-axis angles, realizing a PRISM program on water that matches the far-detector target and acceptance while providing continuous energy–angle coverage.  
By combining beam and atmospheric samples, HK will lift $\delta_{CP}$–mass-ordering degeneracies through matter effects at $E_\nu \sim 3$--$10$~GeV, where DIS dominates and accurate neutrino/antineutrino modeling becomes critical. 

\subsubsection{DUNE}\label{subsec:dune-nd}
The Deep Underground Neutrino Experiment (DUNE) represents the next generation of long-baseline neutrino facilities.  
It will utilize a wide-band beam from Fermilab directed toward a 40~kt liquid-argon far detector located 1300~km away at the Sanford Underground Research Facility (SURF).  
The near-detector (ND) complex comprises three complementary subsystems~\cite{DUNE:2021tad}.  
The ND-LAr is a movable liquid-argon TPC designed to match the far-detector target and technology, providing a direct link between near and far measurements.  
The ND-GAr is a magnetized, high-pressure gaseous-argon TPC with integrated calorimetry, offering low detection thresholds, wide angular coverage, and precise $\nu/\bar{\nu}$ discrimination.  
Finally, the SAND detector is a fixed, on-axis, magnetized spectrometer that continuously monitors the beam and provides multi-target cross-section measurements, including hydrogen-like constraints through material subtraction techniques.  

A defining feature of DUNE’s ND system is its PRISM capability: by moving ND-LAr and ND-GAr off-axis, the detectors can sample distinct narrow-band fluxes whose linear combinations emulate oscillated spectra at the far detector.  
This method substantially reduces model dependence in near-to-far extrapolations and enhances the precision of oscillation measurements.  
The ND program will also carry out dedicated measurements of neutrino–argon interactions, axial and vector form factors, and hadronic energy reconstruction, focusing particularly on controlling missing energy from neutrons and soft hadrons.  
Together, these developments establish DUNE as the most comprehensive experimental platform for precision studies of neutrino interactions and oscillations in the coming decade~\cite{DUNE:2020jqi}.

\subsection{Dedicated Cross-Section Experiments}

A number of dedicated experiments have been designed to measure neutrino–nucleus cross sections with high precision across a wide range of nuclei and kinematics.  
These experiments provide the essential empirical foundation for constraining nuclear models, validating event generators, and reducing systematic uncertainties in oscillation analyses.  
Unlike long-baseline experiments optimized for oscillation measurements, dedicated cross-section programs focus on detailed studies of final-state topologies, hadronic kinematics, and nuclear dependencies.

\subsubsection{MINERvA}
The MINERvA experiment operated in the NuMI beamline at Fermilab from 2009 to 2019, collecting neutrino and antineutrino data on a variety of nuclear targets—He, C, CH, H$_2$O, Fe, and Pb—under identical detector conditions.  
This configuration enabled direct, model-independent comparisons of nuclear effects across targets.  
MINERvA has produced a broad suite of high-statistics measurements, including quasielastic-like and pion-production cross sections with detailed hadronic kinematics, as well as the first target-dependent CC coherent-pion measurement.  
The experiment’s fine granularity allows for precision analyses such as the extraction of the proton axial form factor using $\bar{\nu}$ scattering on hydrocarbon (via carbon subtraction) and low-recoil $\nu_e/\bar{\nu}_e$ cross sections relevant for CP-violation studies.  
At higher $Q^2$, MINERvA explores the shallow and deep inelastic regimes, probing nuclear dependence in the transition to partonic scattering.  
Its extensive dataset, covering multiple targets and interaction modes, is being curated for public release to benchmark future models and event generators~\cite{MINERvA:2017dzh, MINERvA:2018vjb, MINERvA:2018hba, MINERvA:2019rhx, MINERvA:2019kfr, MINERvA:2020zzv}.

\subsubsection{ANNIE}
The Accelerator Neutrino Neutron Interaction Experiment (ANNIE) is a gadolinium-doped water Cherenkov detector located on the Booster Neutrino Beam at Fermilab.  
Its 26~ton total (2.5~ton fiducial) volume is instrumented with large-area picosecond photodetectors (LAPPDs), enabling precision timing and event-topology reconstruction.  
ANNIE’s primary goal is to measure neutron multiplicities from $\nu$–A interactions, directly addressing one of the key uncertainties in long-baseline oscillation analyses: missing energy from undetected neutrons and its impact on reconstructed neutrino energy.  
Operating upstream of SBND, ANNIE enables comparisons between water and argon targets at GeV energies, probing nuclear dependencies in final-state particle production.  
Future upgrades—including the use of water-based liquid scintillator—will improve calorimetric response and hadronic reconstruction, providing a testbed for advanced detector and reconstruction techniques relevant to DUNE and Hyper-Kamiokande~\cite{ANNIE:2019azm, ANNIE:2025vra}.

\subsubsection{NINJA}\label{subsec:ninja}
The Neutrino Interaction Neutron and Jet Analysis (NINJA) experiment employs high-resolution nuclear-emulsion detectors capable of submicron spatial precision and extremely low energy thresholds.  
This technology allows reconstruction of short proton tracks ($p \gtrsim 200$~MeV/$c$) and clean identification of $e^\pm$ versus $\gamma \rightarrow e^+e^-$ conversions, reducing backgrounds in $\nu_e$ samples and isolating 2p–2h topologies.  
The first NINJA physics run deployed a $\sim$250~kg hybrid detector composed of emulsion films interleaved with water, iron, and plastic targets, read out using automated wide-angle scanning systems and paired with a downstream scintillator and magnetized spectrometer (Baby-MIND) for timing and momentum analysis.  
In parallel, studies of heavy-water targets (D$_2$O) aim to separate fundamental $\nu$–N dynamics from nuclear effects through subtraction (D$_2$O–H$_2$O), offering a path toward isolating the underlying elementary amplitudes~\cite{NINJA:2020gbg, NINJA:2022zbi}.

\subsection{Electron--Nucleus Scattering Experiments}\label{subsec:electron-scattering}

High-precision electron–nucleus scattering serves as a cornerstone for understanding the nuclear dynamics underlying neutrino interactions.  
Unlike neutrinos, electrons interact electromagnetically with well-known couplings, and modern accelerators provide monoenergetic, highly collimated beams with precisely known kinematics.  
This enables systematic mapping of nuclear response functions across wide ranges of energy and momentum transfer $(\omega,q)$ in both inclusive and exclusive channels.  
Because the incident beam energy is known event-by-event, electron scattering offers an essential testing ground for neutrino energy reconstruction and model validation under controlled conditions.

Electron–nucleus data directly constrain nucleon momentum distributions, separation energies, and nuclear densities, forming the empirical foundation for spectral-function and mean-field models used in neutrino event generators.  
They also enable detailed studies of final-state interactions that are otherwise challenging to isolate in neutrino measurements.  
By systematically varying the energy and momentum transfer, experiments can disentangle quasielastic, resonance, and deep-inelastic contributions, revealing how nuclear effects evolve across regimes.

The Jefferson Laboratory (JLab) provides the world’s most comprehensive program of high-precision electron scattering, offering beam energies up to 12~GeV and a wide array of nuclear targets ranging from hydrogen and deuterium to carbon, argon, and lead.  
Dedicated Hall~A and Hall~C measurements have mapped the nuclear spectral function, probed short-range correlations, and provided benchmark $(e,e'p)$ and $(e,e')$ datasets that anchor modern neutrino–nucleus models.  
At lower energies, the Mainz Microtron (MAMI) extends precision measurements to the QE and resonance regions up to 1.6~GeV, focusing on nuclear responses relevant to accelerator-based neutrino experiments. Forthcoming facilities such as LDMX and eALBA, together with the MESA energy-recovery accelerator (155~MeV), will expand access to new kinematic windows and precision electroweak observables.  
Collectively, these programs provide indispensable benchmarks for validating theoretical descriptions ensuring a consistent and data-driven understanding of lepton–nucleus interactions~\cite{Ankowski:2022thw, CLAS:2021neh, JeffersonLabHallA:2022cit}.


\section{Summary and Outlook}\label{sec:summary}

Neutrino–nucleus scattering in the medium-energy regime remains one of the most intricate and dynamically evolving frontiers in contemporary nuclear and particle physics.  
The coexistence of quasielastic, resonance, and deep-inelastic processes—together with complex nuclear effects such as Fermi motion, nucleon–nucleon correlations, meson–exchange currents, and final-state interactions—makes it extraordinarily challenging to construct a unified theoretical framework valid across all kinematic regions.  
Despite decades of progress, key uncertainties persist, underscoring the need for continued, coordinated advances in both experiment and theory.

In recent years, remarkable progress has been achieved through the development of high-intensity accelerator neutrino beams and precision detectors.  
Experiments have established a systematic foundation of neutrino–nucleus cross-section data across a wide range of energies, flavors, and nuclear targets.  
These measurements now enable determinations of differential cross sections—crucial for reducing systematic uncertainties in next-generation oscillation programs. Modern short-baseline and dedicated cross-section experiments have been providing a wealth of data, delivering unprecedented detail on exclusive final states and providing essential benchmarks for theoretical models.

On the theoretical side, a new generation of models has emerged that incorporates realistic nuclear ground states, short- and long-range correlations, meson–exchange currents, and in-medium modifications within consistent many-body frameworks.  
At higher momentum transfers, the transition from hadronic to partonic degrees of freedom captured through quark–hadron duality and perturbative QCD - remains challenging to model.  
Meanwhile, advances in lattice QCD and chiral effective field theory ($\chi$EFT) are bridging phenomenology and the fundamental symmetries of Quantum Chromodynamics (QCD), offering systematically improvable and uncertainty-quantified descriptions of nuclear forces and electroweak currents.

This chapter has presented a pedagogical overview of the formalism, mechanisms, and nuclear effects that govern neutrino–nucleus interactions at medium energies.  
It is intended as a self-contained resource for graduate students, postdoctoral researchers, and scientists entering the field, emphasizing both conceptual understanding and practical context.  
By connecting theoretical principles with experimental observables, it highlights how progress in this area depends on the constant dialogue between modeling and measurement.

As neutrino experiments enter the precision era, the close interplay between theory and experiment will be indispensable.  
Each incremental refinement will sharpen our understanding of neutrino–nucleus dynamics and enhance the discovery potential of future experiments.  
Ultimately, this synergy will not only secure the success of DUNE, Hyper-Kamiokande, and other global programs, but will also advance neutrinos as precision tools for exploring the fundamental symmetries and emergent phenomena of the quantum universe.


\begin{ack}[Acknowledgments]%
This work is supported by Fermi Forward Discovery Group, LLC under Contract No. 89243024CSC000002 with the U.S. Department of Energy, Office of Science, Office of High Energy Physics. 
\end{ack}


\seealso{ \\ {\bf 1.} M.~Sajjad~Athar and S.~K.~Singh, \textit{The Physics of Neutrino Interactions}, Cambridge University Press, 2020.} \\
{{\bf 2.} T.~William Donnelly, Joseph A. Formaggio, Barry R. Holstein, Richard G. Milner and Bernd Surrow, \emph{Foundations of Nuclear and Particle Physics}, Cambridge University Press, 2017.}\\
{{\bf 3.} L.~A.~Ruso \textit{et al.}, \textit{Theoretical tools for neutrino scattering: interplay between lattice QCD, EFTs, nuclear physics, phenomenology, and neutrino event generators}, J. Phys. G \textbf{52}, no.4, 043001 (2025) [arXiv:2203.09030 [hep-ph]].} \\
{{\bf 4.} L.~A.~Ruso \textit{et al.}, \textit{NuSTEC White Paper: Status and challenges of neutrino{\textendash}nucleus scattering}, Prog. Part. Nucl. Phys. \textbf{100}, 1-68 (2018) [arXiv:1706.03621 [hep-ph]].}

\bibliographystyle{unsrt}      

\bibliography{reference}

\end{document}